# Cash and Cognition: The Impact of Transfer Timing on Standardized Test Performance and Human Capital[*]

Axel Eizmendi    Germán Reyes

June 2025

## Abstract

This paper shows that the timing of monetary transfers to low-income families affects students' cognitive performance on high-stakes standardized tests. We combine administrative records from the world's largest conditional cash transfer program with college admission exam results of 185,000 high school students from beneficiary families. Exploiting random variation in payment dates, we find that receiving the transfer in the days preceding the exam increases test scores by 0.01 standard deviations relative to receiving it the subsequent week. Question-level analysis reveals that effects are concentrated in final questions and easier questions, suggesting improved cognitive endurance and effort allocation. The impacts are largest for recipients of larger transfers, who experience persistent gains in human capital accumulation: their college enrollment increases by 0.6 percentage points, with higher graduation and formal employment rates seven years later. Our findings show that short-term liquidity constraints during high-stakes events can have long-lasting implications, and suggest opportunities to improve social programs through improved payment scheduling.

[*]Eizmendi (corresponding author): Tufts University (axel.eizmendi_larrinaga@tufts.edu), Reyes: Middlebury College (greyes@middlebury.edu). For helpful discussions and comments, we thank Ned Augenblick, Abhijit Banerjee, Stefano DellaVigna, Kyle Emerick, Armin Falk, Matthew Gudgeon, Johanes Haushofer, Sam Hirshman, Maulik Jagnani, Cynthia Kinnan, John Maluccio, Chris Roth, Dmitry Taubinsky, Ted O'Donoghue, Jonas Radbruch, Frank Schilbach, participants in the Cornell behavioral economics group, participants in the UC Berkeley psychology and economics group, and numerous seminar participants. We also thank Marco Pereira and other members of SEDAP for invaluable help in using the secured data room. Financial support from IZA is gratefully acknowledged.

# 1 Introduction

Cash transfer programs are one of the main anti-poverty policies in developing countries. In Latin America alone, 18 countries operate these programs, reaching over 135 million individuals (Garcia and Saavedra, 2022). Given the scale and central role of cash transfer programs in social protection, understanding how to optimize their effectiveness could benefit millions of households. In this paper, we focus on payment timing and examine how it affects beneficiaries' cognitive performance—a fundamental determinant of decision-making and predictor of life success (e.g., Heckman et al., 2006; Kuncel and Hezlett, 2007; Hanushek and Woessmann, 2008; Burks et al., 2009; Dohmen et al., 2010; Benjamin et al., 2013; Fe et al., 2022).

To investigate this, we study Brazil's *Bolsa Familia Program* (BFP)—the world's largest conditional cash transfer (CCT) program by coverage, which serves over 13 million households. BFP provides monthly transfers to low-income households, granting on average $93 per month, which represents approximately 40 percent of beneficiaries' monthly income. To measure cognition, we combine BFP administrative records with data from 185,000 high school students from beneficiary families who took a high-stakes test used for college admissions called ENEM.

Our identification strategy exploits exogenous variation in payment dates. The day of the month a household receives its BFP payment is determined by the last digit of a randomly-generated beneficiary identifier. In our sample, approximately 60 percent of households receive their monthly BFP transfer before the ENEM exam ("early recipient"), while the remainder receive it the week after ("late recipient"). Over 90 percent of early recipients withdraw their benefits prior to the ENEM exam, creating quasi-experimental variation in household liquidity at the time of the exam.

Our main result is that receiving the transfer before the exam improves test performance. Early recipients score approximately one percent of a standard deviation higher than late recipients. We find positive effects across all exam subjects, with the largest effects in language arts and the essay. The effects are concentrated in the middle of the score distribution and are driven by large-transfer recipients—those in the top third of transfer size—with limited effects for recipients of smaller transfers. Importantly, these effects are not driven by changes in test-taker composition, as early recipients have the same probability of attending the exam as late recipients.

To investigate potential mechanisms, we analyze question-level data using two sources of variation. First, we exploit the random assignment of test booklets to students, where all booklets have the same set of questions but in different positions. Second, we leverage natural variation in question difficulty. We find that the effects are concentrated in the final questions of the exam, consistent with increased cognitive endurance (Sievertsen et al., 2016; Mani et al., 2013; Reyes, 2023; Brown et al., 2024). We also find larger effects on easier exam questions, but negative effects on the most challenging questions that exceed students' ability. This pattern suggests that early recipients more strategically allocate their time and cognitive resources across exam questions.

Next, we assess whether these performance gains have persistent consequences for human capital accumulation. The ENEM exam plays a central role in Brazil's higher education system, determining access to scholarships, student loans, and university admissions for over five million test-takers annually (Machado and Szerman, 2021). Given the exam's scale, even modest increases in ENEM scores could significantly affect students' college trajectories. To examine potential effects on college outcomes, we link ENEM records to Brazil's higher education census, focusing on large-transfer recipients—the group where we found effects on test scores.

We find that early transfer payments have positive effects on human capital accumulation. Among large-transfer recipients, the college attendance rate increases by 0.6 percentage points (or 3.9 percent relative to a baseline of 15.3 percent). We provide suggestive evidence that increased access to government scholarships serves as a mechanism behind the higher enrollment rates, as early recipients in this group are more likely to meet the minimum test score requirements for scholarship eligibility. Importantly, these high-benefit students succeed academically despite having lower academic preparation and limited financial resources. We analyze degree completion rates up to seven years after the ENEM exam and find that seven-year graduation rates increase by 0.8 percentage points (or six percent relative to a baseline of 13 percent) for this group.

A natural question is whether the improved college outcomes translate into better labor market outcomes. To assess this, we link ENEM test-takers to employer-employee matched data covering the universe of Brazil's formal-sector workers. Since we observe students for up to nine years after the ENEM exam, our analysis captures early-career outcomes. For large-transfer recipients, receiving transfers before the exam increases formal employment rates by 0.9 percentage points (or 3.1 percent relative to late recipients). We find no significant effects on earnings or wages during this period, though this null result may reflect that many students are still completing their degrees or early in their career trajectories.

Our results demonstrate that liquidity constraints, even when short-lived, can have persistent effects when they coincide with high-stakes events such as college entrance exams. Moreover, the concentration of effects among large-transfer recipients—who have lower household income—suggests that the magnitude of liquidity relief matters for cognitive performance. Our finding points to a low-cost opportunity to improve the effectiveness of cash transfer programs through optimized payment timing. Program administrators could improve participants' outcomes by aligning transfer schedules with key events in beneficiaries' lives, without requiring changes to benefit levels or eligibility criteria.

This paper contributes to the literature on CCTs and educational outcomes.[1] Previous work finds positive impacts of CCTs on school enrollment, attendance, and dropout rates (e.g., Glewwe and Kassouf, 2012; García and Saavedra, 2017). However, evidence on CCTs' effects on test scores is mixed (Millán et al., 2019). A key empirical challenge is that CCT benefits are often conditional on school attendance, which changes the composition of test takers (Garcia and Saavedra, 2022). Most existing work thus relies on strong identification assumptions to address selection into test-taking.[2] We circumvent this challenge by exploiting variation in *when* households receive their transfer, which allows us to measure impacts on test scores under weaker identification assumptions.[3] We also show that these test score effects translate into meaningful differences in college enrollment and completion.

We also contribute to the literature on income shocks and cognitive performance. Recent empirical work shows that poverty can directly impair cognitive functioning (Mani et al., 2013; Schilbach et al., 2016; Dean et al., 2019), that income shocks affect performance on cognitive tests (Lichand and Mani, 2020; Kansikas et al., 2023), and that financial concerns impact task performance (Duquennois, 2022; Kaur et al., 2025).[4] Our paper advances this literature in three ways. First, we examine effects on household dependents rather than direct transfer recipients. Second, we study impacts on a high-stakes test rather than labo-ratory measures of cognition.[5] Third, because we study an existing nationwide policy, our estimates have direct policy relevance and benefit from a substantially larger sample than previous studies.[6]

---

[1] For overviews of this literature, see Fiszbein and Schady (2009); Baird et al. (2014); Millán et al. (2019); Garcia and Saavedra (2022).

[2] See for example Behrman et al. (2000); Garcia and Hill (2010); Báez and Camacho (2011). A small number of studies test randomly-selected children at home to avoid selection (e.g., Evans et al., 2014; Barrera-Osorio and Filmer, 2016; Paxson and Schady, 2010).

[3] By focusing on immediate effects, our design rules out many indirect channels through which cash transfers can impact cognition. For instance, one of the conditions to remain enrolled in BFP is for school-age children to be enrolled in school and attend. Thus, BFP could have long-term impacts on cognition partly due to their impact on schooling (Hansen et al., 2004; Brown et al., 2024).

[4] Though other work fails to find impacts of income shocks on cognition (Carvalho et al., 2016; Fehr et al., 2022), and recent reviews suggest the overall evidence is mixed (De Bruijn and Antonides, 2022; Haushofer and Salicath, 2023).

Finally, we contribute to the literature on the optimal design of cash transfer programs. Previous work has analyzed the impact of conditioning transfers on household behaviors, "labeling" the intended use of benefits, and comparing in-kind to cash transfers (e.g., Baird et al., 2011; Cunha, 2014; Attanasio et al., 2015; Benhassine et al., 2015; Cunha et al., 2019). The *timing* of transfers has received less attention as a design feature, even though evidence from other social programs suggests that payment timing can substantially affect household behavior (e.g., Hastings and Washington, 2010; Goldin et al., 2022; Bond et al., 2022).[7] We provide novel evidence that transfer timing can impact longer-run outcomes like college attendance and graduation, offering insights into how improved program design can enhance human capital accumulation.

## 2 Institutional Context and Data

### 2.1 Cash Transfer Program

The *Bolsa Família Program* (BFP) is a conditional cash transfer (CCT) program that provides monthly benefits to low-income families. BFP covers about 14 million households in Brazil, reaching over 50 million individuals (approximately 25 percent of the country's population), making it the largest CCT program worldwide (Brollo et al., 2020).

BFP benefits are determined by household composition and income through a two-tier structure. Households below the extreme poverty line receive both an unconditional transfer and conditional payments tied to children's school attendance and health checkups. Households between the extreme poverty and poverty lines are eligible only for the conditional component.[8] In 2013, the year we analyze, the average beneficiary family

---

[5]Incentives are important for measuring cognitive performance because in their absence, cognitive tests capture not only cognitive ability but also other performance determinants such as intrinsic motivation (Wise and DeMars, 2005; Duckworth et al., 2011; Segal, 2012; Finn, 2015; Gneezy et al., 2019).

[6]A related literature on "payday effects" documents significant intra-month declines in liquidity and consumption as households approach their next paycheck, social security benefit, or food stamps benefit (Stephens Jr, 2003; Shapiro, 2005; Mastrobuoni and Weinberg, 2009; Hastings and Washington, 2010). We show in a developing country context that these effects can have significant long-term consequences when they coincide with high-stakes events.

[7]Notable exceptions include Barrera-Osorio et al. (2011) and Kansikas et al. (2023).

[8]The conditional benefits include payments for pregnant women and nursing mothers who meet health monitoring requirements, school-age children under 17 who maintain minimum attendance, and children

received a monthly payment of $93, representing approximately two-thirds of their average monthly income of $138.

To enroll in BFP, households must register in the Social Programs Registry (*Cadastro Único*). During registration, each household is assigned a randomly generated eleven-digit identification number called *NIS*. The disbursement schedule of the benefits each month is directly linked to the last digit of the NIS number. Households whose NIS ends with the digit '1' are the first ones to receive the transfer, typically on the third Monday of the month. Thereafter, transfers are made sequentially, with each subsequent day serving those whose NIS ends with the next sequential digit, and with no disbursements occurring over the weekends and public holidays. Figure 1 shows the October 2013 payment schedule. Our identification strategy, described in Section 3, exploits the variation in payment timing induced by the NIS last digit.

### 2.2  ENEM Exam

The High School Assessment Exam (*Exame Nacional do Ensino Médio*, or ENEM) is a high-stakes exam that plays a central role in Brazil's higher education system. Most universities in Brazil, including those in the centralized admission system, use ENEM scores as their sole admission criterion. Application components common in other countries, such as extracurricular activities, interviews, or motivation essays, play no role in these institutions. ENEM scores also determine eligibility for government scholarships through ProUni (*Programa Universidade para Todos*) and student loans through FIES (*Fundo de Financiamento ao Estudante do Ensino Superior*). In 2014, only 16 percent of private college enrollees reported using ENEM scores for admission, highlighting the exam's particular importance for access to the public university system.

The ENEM exam consists of an essay and 180 multiple-choice questions across four subjects: mathematics, natural sciences, social sciences, and language arts. Each exam subject is graded based on Item Response Theory (IRT). Testing occurs once a year over two consecutive weekend days. Test-takers are randomly assigned an exam booklet that contains the same set of questions but in different orders within each subject area. Over five million individuals enroll to take the ENEM exam every year, making it the second largest college admission exam in the world.

---

under seven who follow vaccination schedules (Glewwe and Kassouf, 2012).

## 2.3 Data

We combine several administrative datasets. First, we use the 2013 ENEM exam records, which contain both student- and question-level information. The student-level data include test scores and demographic characteristics such as age, gender, race, and family income. The question-level data provide information about the subject and skill being tested, as well as measures of question difficulty.

We identify students from BFP recipient households by linking ENEM records to the Registry of Social Programs and BFP payment sheets from 2013. The Registry identifies households receiving BFP. The payment sheets contain information on transfer size, withdrawal dates, and beneficiaries' NIS numbers, which determine the day households receive their monthly transfer.

To examine longer-run outcomes, we link these records to two additional datasets. The Census of Higher Education (2014-2021) provides information on college enrollment, field of study, degree progress, and graduation outcomes. The RAIS employee-employer matched data (2014-2022) contains earnings, occupation, and industry information for all formal-sector workers in Brazil.

## 2.4 Sample, Summary Statistics, and Balance

We focus on high school seniors from BFP recipient households who enrolled to take the 2013 ENEM. To construct our sample we implement several sample restrictions. First, we exclude individuals who take the exam in their junior year, as these students typically do so for practice. Second, we exclude individuals who take the exam after dropping out of high school, who typically take the ENEM to obtain a high school equivalency certificate. Third, we exclude individuals who take the exam after graduating from high school, who often take the exam to enroll in a college degree or switch from college degrees.[9] Fourth, we exclude students whose families are not enrolled in BFP. Finally, we exclude a small fraction of students from households that were enrolled in BFP but did not receive transfers during the exam month, typically due to ineligibility or non-compliance with program requirements. These restrictions yield a sample of approximately 185,000 students.

Table 1 presents summary statistics for our sample and comparison groups. Column 1 shows characteristics of all high school seniors who registered for the ENEM exam, while columns 2–3 and 4–5 compare students by their enrollment status in the Social Programs

---

[9]To ensure we capture potential new entrants into higher education, we exclude students who appear in the higher education census as enrolled in college during or prior to taking the ENEM exam.

Registry and BFP, respectively. Our analysis sample, shown in column 4, consists of students who are older (20.5 years versus 18.9 in the full sample), more likely to be female (65.2 percent versus 58.8 percent), and more likely to identify as Black or Brown (72.4 percent versus 53.5 percent overall). Students in our sample come from disadvantaged backgrounds, with only 1.8 percent attending private high school (compared to 18.6 percent overall) and 2.0 percent having college-educated mothers (versus 16.2 percent in the full sample). Among test-takers, BFP beneficiaries score approximately 0.6 standard deviations below non-beneficiaries across all exam subjects.

The sample is balanced on observable characteristics across payment timing groups. Table 2 compares mean characteristics of students whose households receive BFP transfers before versus after the ENEM exam (columns 1 and 2). The differences between groups are small in magnitude (column 3), and for most variables, we cannot reject equality at conventional significance levels. We also find a balanced sample when examining student characteristics across all ten possible NIS last digits (Appendix Table A1). Consistent with this, the last digit of household NIS numbers is uniformly distributed across students (Appendix Figure A1).

## 3 Research Design

### 3.1 Regression model

Our empirical strategy compares test scores between students whose households received their BFP transfer before versus after the ENEM exam. As described in Section 2, this variation stems from the last digit of households' randomly-assigned NIS numbers, which determines payment timing. We estimate the following baseline specification:

$$Y_i = \alpha + \beta \text{BFPBeforeENEM}_i + \varepsilon_i, \tag{1}$$

where $Y_i$ is student $i$'s outcome and $\text{BFPBeforeENEM}_i$ is an indicator that equals one if the last NIS digit of $i$'s family allows them to receive the transfer before the exam. Standard errors are clustered at the student level.

The coefficient $\beta$ measures the causal effect of having access to BFP funds before versus after the exam. We interpret this parameter as the effect of alleviating end-of-cycle financial constraints—the liquidity shortages that households typically face in the days before their next transfer payment. Rejecting the null hypothesis that $\beta = 0$ would provide evidence that households face binding end-of-cycle constraints that affect cognitive performance.

## 3.2 Main Outcomes

We examine three categories of outcomes (see Appendix B.1 for details). First, we analyze test performance (see Appendix C for a conceptual framework that connects cash transfers to cognitive functions and test performance). We use the IRT-based scores provided by the testing agency, standardized to have mean zero and standard deviation one in our sample. We analyze both overall performance (measured as the student's average IRT-score across all subjects) and performance in each component: mathematics, natural sciences, social sciences, language arts, and the essay.

To assess longer-run impacts, we examine both college and labor market outcomes. For college outcomes, we measure enrollment (appearing in the higher education census one year after ENEM), college quality (measured by graduates' mean earnings in administrative data), persistence (remaining enrolled in years 1-3), and degree completion (graduating within seven years of ENEM). For labor market outcomes up to nine years after the ENEM, we analyze formal employment (appearing in the employer-employee matched data), monthly earnings, and hourly wages. Earnings and wages are inflation-adjusted and expressed in $2023.

## 3.3 Transfer Payment Timing and Withdrawal Patterns

We begin by examining the timing of household benefit withdrawals relative to BFP disbursement dates. Figure 2 plots the cumulative distribution function (CDF) of withdrawals relative to scheduled payment dates. Since payment dates vary across households based on their NIS number, we normalize each household's payment date to zero. The *x*-axis represents days relative to the payment date, while the *y*-axis axis shows the cumulative share of households that have withdrawn their benefits.

Withdrawal patterns indicate that households face binding liquidity constraints and value immediate access to funds. We observe virtually no withdrawal activity before the scheduled payment date. On the payment date itself, approximately 75 percent of households withdraw their benefits, and this share rises to over 85 percent within three days. This rapid withdrawal behavior means that the vast majority of households scheduled to receive transfers before the exam had accessed their funds prior to the exam weekend (Appendix Figure A2).

## 4 The Impact of Monetary Transfers on Exam Performance

This section examines the impact of the timing of BFP cash transfers on students' exam performance. We begin by assessing whether receiving BFP early affects the likelihood of attending the ENEM. Then, we turn to our main analysis and investigate the impact of receiving the cash transfer on the overall test and across different test subjects. We conclude the section by studying heterogeneity by transfer size and student characteristics.

### 4.1 Exam Attendance

We begin by studying selection into taking the ENEM exam. Receiving the cash transfer in the days preceding the exam could impact the individual's decision to take the exam. Moreover, if the students induced to take the exam have systematically different cognitive abilities, this effect could lead to test-score differences between early- and late-receivers, irrespective of any cognition effects.

To measure the impact of BFP on the probability of attending the exam, we define indicator variables for whether students were present on both exam days, at least one day, and respectively on the first and second day. Table 3 presents estimates of equation (1) using these attendance indicators as outcomes. Appendix Figure A3 plots the proportion of BFP beneficiaries who attended both exam days (blue circles) and at least one day (red triangles) against the days elapsed since receiving the transfer.

We find no evidence that transfer timing affects exam attendance. The impact of BFP on attendance for both days is of $-0.1$ percentage points (on a baseline of $75.8$ percent), and the impact on attending at least one day is of $-0.2$ percentage points (on a baseline of $78.5$ percent). These effects are small and statistically insignificant. Results are similar for attendance on the first day (column 3) and second day (column 4). Consistent with the regression results, Appendix Figure A3 confirms that attendance rates do not systematically vary with the timing of household transfer receipt.

### 4.2 Test Scores

We begin our main analysis by examining how test performance varies with cash transfer disbursement dates. Figure 3 plots average test scores against the timing of BFP disbursement relative to the exam date.

Students whose households receive transfers before the exam tend to perform better than those who receive transfers afterwards. For instance, students receiving transfers five days before the exam score 0.0027 standard deviations above the mean (s.e. = 0.0025 SD), while those receiving transfers four days after score 0.0061 standard deviations below the mean (s.e. = 0.0031 SD). Test scores are similar among students who receive transfers at different times before the exam (henceforth "early recipients"), and likewise among those who receive transfers at different times after the exam (henceforth late recipients"). Given this pattern, we proceed by analyzing the difference in outcomes between these two groups rather than examining day-by-day variation in transfer timing.

Table 4 presents estimates of $\beta$ from equation (1), which captures the mean score difference between early and late recipients. Column 1 reports results for the average score across all subjects, while columns 2–6 present estimates separately for each multiple-choice subject and the essay. For visual evidence of these effects, Appendix Figure 4 plots mean test scores for early and late recipients by subject. Appendix Table A12 examines the impact of early BFP receipt on the probability of exceeding different score thresholds on the original ENEM scale, which has a mean of 500 and standard deviation of 100.

Receiving the cash transfer in the days preceding the exam increases test scores. On average, receiving BFP early improves test scores by $\beta = 0.009$ SD ($p < 0.05$), as shown in column 1. This figure represents a 1.69 percent increase relative to the late-recipients' average score. Notably, the performance of early-recipients is better on all subjects tested (Appendix Figure 4). The impact of BFP is larger for the language arts test ($\beta = 0.014$ SD, $p < 0.01$) and the essay ($\beta = 0.010$, $p < 0.10$), with test score increases of approximately 2.5 percent relative to the late recipients mean (columns 4 and 6). While we also estimate positive impacts on the sciences and math (columns 2, 3, and 5), these effects are not statistically significant. The effects are concentrated in the middle of the test score distribution, with early BFP receipt significantly increasing the probability of scoring above 400–500 points but having no impact on the likelihood of achieving either very low or very high scores (Appendix Table A12). The results are robust to controlling for socioeconomic and demographic characteristics (Appendix Table A2).

A potential threat to identification is that other government programs might disburse benefits according to households' last NIS digits. If the timing of these disbursements coincides with BFP payments, it could confound our estimates. We address this potential concern by examining students from households with an NIS identifier who do not receive BFP.[10] This placebo group allows us to isolate any effects of non-BFP programs that use NIS digits for payment timing. Regression estimates for this group, reported in Appendix Table A4, are substantially smaller than our baseline estimates and statistically insignificant.[11]

---

[10]As detailed in Section 3, households receive an NIS when they enroll in the Social Registry Program, which is required for all government benefits (including, but not limited to, BFP).

Our estimates indicate that early receipt of BFP improves exam performance by approximately one percent of a standard deviation, representing a two percent increase relative to late recipients' mean scores. This effect is comparable to increasing teacher value-added by 0.06 standard deviations (Chetty et al., 2014), reducing air pollution during testing by 0.20 standard deviations (Ebenstein et al., 2016), lowering ambient pollen levels by 0.32 standard deviations (Bensnes, 2016), or decreasing testing room temperature by one degree Fahrenheit (Park, 2020).[12]

### 4.3 Heterogeneous Treatment Effects

**4.3.1 Transfer Size.** Understanding whether impacts vary with transfer size can inform the optimal design of benefit levels in cash transfer programs. In our sample, we observe substantial variation in transfer sizes. The average household receives $94 per month, with a standard deviation of $49. To analyze heterogeneity, we divide students into terciles based on their household's monthly benefit. Average monthly transfers are $51, $92, and $175 in the bottom, middle, and top terciles, respectively. Households in the top tercile are substantially more dependent on BFP as they are larger and have lower income (Appendix Table A5). Table 5 presents estimates of equation (1) separately for each tercile.

Test score effects are concentrated among recipients of the largest transfers. Among students in the top tercile of transfer size (henceforth "large-transfer recipients"), receiving BFP early improves average test scores by $\beta$ = 0.02 standard deviations ($p$ < 0.01), representing a 3.3 percent increase relative to the mean score of late recipients in this group (Panel A, column 1). The effects are positive across all subjects, with the largest gains appearing in language arts ($\beta$ = 0.026 SD, $p$ < 0.05) and the essay ($\beta$ = 0.025 SD, $p$ < 0.05). In contrast, we find substantially smaller and generally insignificant effects for students in the middle and bottom terciles of transfer size (Panels B and C). These patterns persist when we control for socioeconomic and demographic characteristics (Appendix Table A6).

---

[11]This finding aligns with information from Ministry of Citizenship staff—the agency in charge of managing BFP—who confirmed that no other programs used NIS digits to determine payment schedules during this period.

[12]Another way to benchmark our effect size is to estimate the public funds that would be required to achieve a corresponding test score increase. In a meta-analysis of the impact of public school spending on student test scores, Jackson and Mackevicius (2023) finds that increasing spending by $1,000 per student for four years leads to a test-score increase of 0.031 SD. Extrapolating this US-based estimate to Brazil and assuming that the impact of public funds scales linearly, our estimated increase of 0.009 SD (Table 4, column 1) would require approximately $260 per student for four years (or about $115 million, given the number of students who are early-recipients).

**4.3.2 Student characteristics** Understanding whether specific types of students particularly benefit from the cash transfer is important for program targeting (e.g., Berger et al., 2000; Manski, 2004; Dehejia, 2005; Kitagawa and Tetenov, 2018; Athey et al., 2023). We analyze student gender, race, employment status, and mother's education. For each characteristic, we estimate an augmented version of equation (1) that includes both an indicator for the characteristic and its interaction with early BFP receipt. Appendix Tables A6–A8 presents estimates of the baseline effect and interaction terms.

We find heterogeneity only in the education level of test-takers' mothers. The impact of BFP on test scores is $0.057$ SD larger for students with college-educated mothers ($p < 0.10$), as shown in Appendix Table A8, column 2. This pattern may reflect that college-educated mothers have higher perceived returns to their children's education (Boneva and Rauh, 2018) and consequently direct more of the cash transfer toward educational investments. Other student and household characteristics do not significantly moderate BFP's impact. For instance, while students who work 20 hours per week experience a $0.006$ SD smaller effect than non-working students, this difference is not statistically significant (column 1). We find similarly negligible differences between female and male students and between white and non-white students (columns 3 and 4).

## 4.4 Mechanisms

A unique feature of the ENEM is that it contains detailed question-level information, which we leverage to investigate the mechanisms underlying improved test performance. We analyze heterogeneity along two dimensions of exam questions: their position within the test and their difficulty level. Given that the test score effects are concentrated among recipients of larger transfers, we focus our analysis on this subgroup.

**4.4.1 Question position and cognitive fatigue.** First, performance gains could stem from an increased ability to endure cognitive fatigue as the exam progresses. To test this mechanism, we examine whether early BFP receipt has larger effects on questions that appear later in the exam, when fatigue is most likely to set in. We leverage the random assignment of exam booklets to generate exogenous variation in question position and estimate the following regression at the student-question level:

$$\text{Correct}_{ij} = \alpha + \beta \text{BFPBeforeENEM}_i + \varphi \text{AmongLastTen}_{ij}$$
$$+ \gamma (\text{BFPBeforeENEM}_i \times \text{AmongLastTen}_{ij}) + \varepsilon_{ij}, \quad (2)$$

where $\text{Correct}_{ij}$ equals one if student $i$ correctly answered question $j$, and $\text{AmongLastTen}_{ij}$ equals one if question $j$ is among the last ten questions of either test day. Appendix Table A9 presents estimates of $\gamma$, which captures the differential impact of early BFP receipt on performance in the final exam questions. For visual evidence of these effects, we construct 180 indicator variables for correct responses at each position and estimate equation (1) using each as an outcome. Appendix Figure A4 plots these position-specific treatment effects against question order.

Early BFP receipt has significantly larger effects on performance in later questions. The impact on the probability of correctly answering a question is $\gamma = 0.2$ percentage points higher ($p < 0.05$) for questions appearing at the end of either test day (column 1). This effect is primarily driven by the second day of testing, where the differential impact on the last ten questions is $\gamma = 0.3$ percentage points, compared to $\gamma = 0.1$ percentage points on the first day (columns 2 and 3). Consistent with these regression results, Appendix Figure A4 reveals a clear clustering of positive effects in the final portions of the exam, suggesting that early transfer receipt particularly improves performance when cognitive fatigue is most likely to set in.

**4.4.2 Question difficulty and effort allocation.** Second, we examine whether early BFP receipt affects how students allocate effort across questions of varying difficulty. We classify questions into percentile ranges based on the out-of-sample share of correct responses. Questions in the bottom quartile of difficulty are particularly challenging for BFP beneficiaries—the share of correct responses among late recipients (13–16 percent) falls below what would be expected from random guessing (20 percent). In contrast, questions in the $90^{\text{th}}$–$100^{\text{th}}$ percentile range are more within students' ability, with a baseline correct response rate of 46.3 percent. Appendix Table A10 presents estimates of equation (1) using the share of correct responses in each difficulty category as outcomes.

Early BFP receipt increases performance on easier questions but reduces performance on the most challenging ones. The share of correct responses to the easiest questions ($90^{\text{th}}$–$100^{\text{th}}$ difficulty percentile) increases by 0.4 percentage points ($p < 0.05$), while the share of correct responses to the hardest questions falls by 0.14–0.16 percentage points (columns 1–2). We find no significant effects on questions of medium difficulty (columns 3–4) or questions in the 75$^{\text{th}}$–90$^{\text{th}}$ difficulty percentile. This pattern suggests that early BFP receipt leads students to reallocate effort toward questions where they have a higher probability of success.

## 5 Longer-Run Outcomes

This section examines whether the test score gains from early BFP receipt translated into improvements in longer-run educational and labor-market outcomes.

### 5.1 College Enrollment, Quality, and Graduation

Early BFP receipt increased college enrollment rates, with effects concentrated among recipients of larger transfers (Table 6). For the full sample, receiving BFP before the exam increased college enrollment by $\beta = 0.2$ percentage points, though this estimate is not statistically significant (Panel A, column 1). However, among large-transfer recipients, early receipt increased enrollment by $\beta = 0.6$ percentage points ($p < 0.10$), representing a 3.9 percent increase relative to the late-recipients' baseline enrollment rate of 15.3 percent (Panel B, column 1). Importantly, early BFP receipt led to an increase in college quality—as measured by expected graduate earnings—with students in the top transfer tercile attending programs where graduates earn $\beta = \$5.4$ more per month (or 1.6 percent more) on average than late recipients ($p < 0.05$, column 4).

The increases in initial college enrollment led to sustained gains in educational attainment among large-transfer recipients (Table 6). Early BFP recipients in this group had a $\beta = 0.8$ percentage points higher three-year persistence rate after the ENEM exam ($p < 0.05$), which represents a 4.1 percent increase relative to the late-recipient mean of 19.6 percent (Panel B, column 5). These persistence gains translated into higher degree completion rates; students whose households received transfers early were $\beta = 0.8$ percentage points ($p < 0.05$) more likely to graduate within seven years of the exam (Panel B, column 6). The magnitude of these effects aligns with our enrollment findings, suggesting that the marginal students induced to attend college by early transfer receipt succeeded academically. We find no significant effects on persistence or graduation rates in the full sample or among recipients of smaller transfers (Panels A, C, and D).

## 5.2 Mechanisms for College Access

To understand how higher test scores translate into increased college enrollment, we examine two main channels through which early BFP receipt affects college access.

### 5.2.1 Admission to Selective Programs.

First, we examine whether higher test scores increased admission chances at selective universities. Better scores could mechanically increase admission probabilities if they push students above program admission cutoffs. To assess this, we analyze enrollment patterns at institutions that use ENEM scores for admissions, focusing particularly on public universities where admissions are highly competitive and exclusively determined by ENEM performance. We do not find evidence that increased college enrollment stems from students clearing critical admission thresholds. We find null effects on enrollment at both public colleges (Table 6, column 3), which comprise the majority of institutions that use the ENEM for admissions, and selective private colleges that use ENEM scores (Table 7, column 1). These null results are consistent with the highly competitive nature of public university admissions, where cutoff scores substantially exceed the performance of BFP beneficiaries.[13] Since the effects on test scores are concentrated in the middle of the score distribution (Appendix Table A12), these improvements are unlikely to affect admission prospects at selective institutions.

### 5.2.2 Access to Financial Aid.

Second, we examine whether higher test scores increased access to financial aid. Higher scores could increase access to merit-based government scholarships through *ProUni*, a federal program that provides full and partial tuition scholarships to low-income students at private universities. These scholarships are particularly important for BFP recipient households who typically attend non-selective private colleges, as they are allocated competitively based on ENEM scores. To be eligible for these scholarships, applicants must achieve an average ENEM score above 450 points across all subjects and a non-zero score on the essay. In our sample, 51.2 percent of late recipients satisfy the average score requirement and 96.1 percent satisfy the essay requirement, suggesting that the score threshold may be a binding constraint for many students.

Early BFP receipt increases students' likelihood of meeting *ProUni* eligibility requirements (Table 7, columns 2–3). On average, early recipients are 0.6 percentage points more likely to exceed the minimum score threshold ($p < 0.05$), representing a 1.1 percent increase relative to late recipients. They are also 0.02 percentage points more likely to meet the essay requirement, though this effect is not statistically significant. These effects are larger and more precisely estimated among recipients of larger transfers (Panel B), while we find positive but statistically insignificant effects for other students (Panels C and D).

---

[13]Appendix Figure A5 illustrates this by comparing public university admission thresholds from the centralized admission system in 2017 (earliest year for which we have access to this data) to the distribution of low-income students' ENEM scores.

The improved eligibility translates into suggestive evidence of increased scholarship receipt, particularly among large-transfer recipients. Among late recipients, 8.3 percent receive full scholarships and 6.9 percent receive partial scholarships. While we find no significant effects in the full sample (Panel A), students receiving larger transfers show meaningful increases in scholarship access. Early receipt increases their likelihood of receiving a full scholarship by 0.3 percentage points (4.1 percent) and a partial scholarship by 0.5 percentage points (8.3 percent). Though these estimates are not statistically significant due to smaller sample sizes, their magnitudes align with the increases in eligibility and overall college enrollment.

## 5.3 Labor Market Outcomes

We next examine potential impacts on labor market outcomes. The Brazilian labor market is characterized by high rates of informality. In 2023, about 39 percent of workers were employed in the informal sector. Formal-sector jobs tend to offer better benefits, such as health insurance, paid sick leave, unemployment insurance, and retirement benefits, but access to these jobs is particularly limited for individuals with low educational attainment (Haanwinckel and Rodrigo, 2017). Given the positive effects on college outcomes documented in Section 5, we test whether the timing of BFP affects workers' success in securing formal employment and their subsequent earnings.

Early BFP receipt increases formal employment rates but has no detectable effect on earnings. Table 8 shows null average effects across all labor market outcomes in the full sample (Panel A). However, among large-transfer beneficiaries, receiving BFP before the exam increases the probability of formal employment by 0.9 percentage points ($p < 0.05$), representing a 3.1 percent increase relative to the late-recipient mean of 28.8 percent (Panel B). Despite the higher formal employment rate, we find precisely estimated zero effects on earnings and wages, potentially because many individuals are still completing their degrees or early in their career trajectories.

# 6  Discussion

This paper shows that liquidity constraints can have persistent effects on economic mobility when they coincide with high-stakes events. We show that receiving a cash transfer shortly before a college entrance exam improved students' cognitive performance, particularly among recipients of larger transfers. These test score gains translated into meaningful differences in educational attainment, with early recipients more likely to both enroll in and complete college degrees. The improved human capital accumulation ultimately led to better labor market prospects, with early recipients of larger transfers becoming more likely to secure formal-sector employment.

Our results suggest a previously unexplored mechanism by which persistence in low-income status can perpetuate. A precarious financial situation directly impairs performance on cognitive tasks, which subsequently influences access to education and human capital accumulation. Human capital, in turn, shapes future financial outcomes. By highlighting this linkage, our findings suggest that even temporary relaxation of liquidity constraints can meaningfully impact economic mobility when timed to coincide with high-stakes events. Importantly, if households are unaware of these "cognition effects," information-provision policies that encourage households to shift important decisions away from low-income periods could be welfare-improving even if they do not change the present discounted value of the households' income.

Finally, our findings suggest that the timing of cash transfer programs can be an effective policy tool to improve the outcomes of low-income households. There are many critical one-time events where maximizing performance or making the right decision can have long-lasting consequences (e.g., job interviews, choosing the right savings plan or insurance provider, etc.). Although personalizing transfer timing to each household's unique needs might be infeasible, offering households flexibility in selecting payment dates within a window could be a practical approach. Our results suggest that such simple modifications to existing programs could generate improvements without requiring changes to benefit levels or eligibility criteria.

# Figures and Tables

Figure 1: Timing of Bolsa Familia disbursements and ENEM exam during October 2013

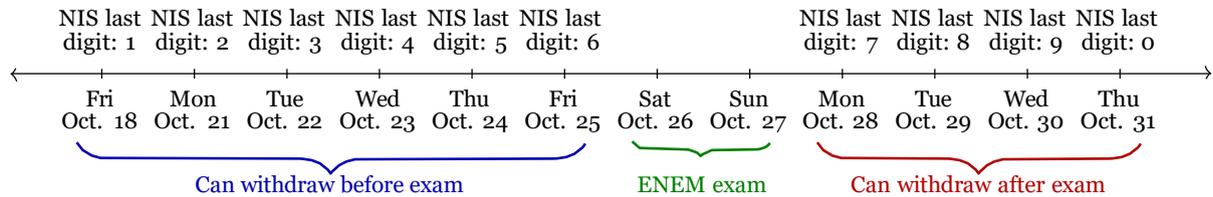

*Notes*: This figure shows the BFP transfer payment schedule for October 2013. Payments are sequentially disbursed across ten consecutive weekdays (excluding public holidays), typically starting on the third Monday of each month. The payment date is determined by the last digit of each beneficiary's randomly assigned NIS number. NIS numbers ending in '1' receive payment on the first day, followed by sequential payments to subsequent digits on following days.

Figure 2: Household withdrawal behavior following BFP payment date

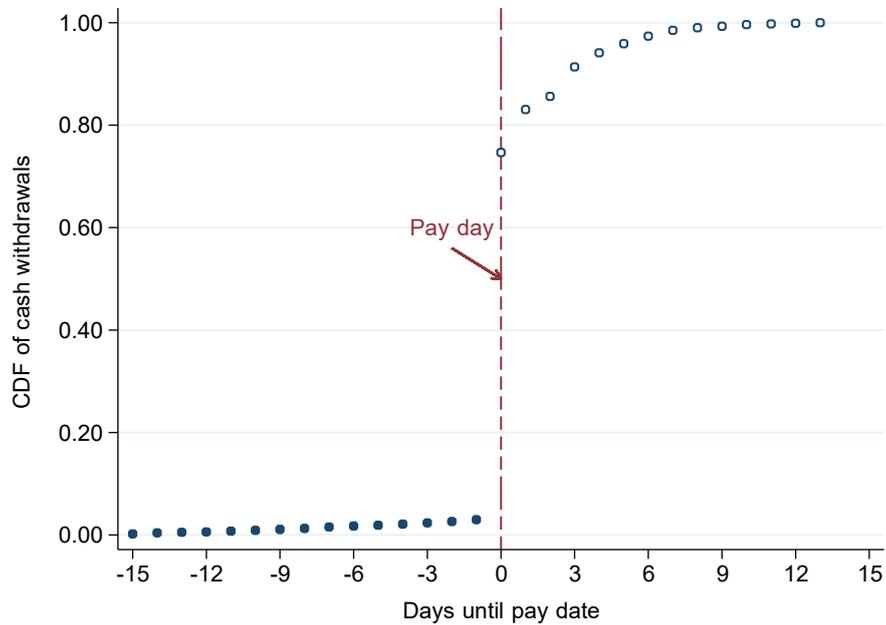

*Notes*: This figure shows the cumulative distribution function (CDF) of cash withdrawals relative to households' scheduled BFP payment dates. The *x*-axis shows days relative to the scheduled payment date (normalized to 0), while the *y*-axis shows the fraction of households that have withdrawn their benefits by each day. The vertical dashed line indicates the scheduled payment date.

## Figure 3: Average ENEM score by days since BFP cash transfer

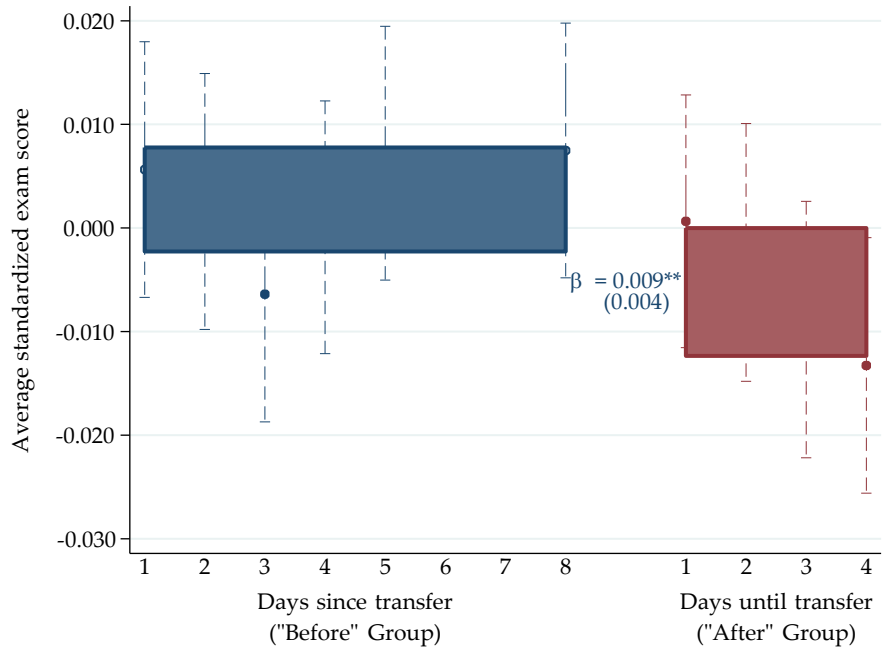

*Notes*: This figure shows the average standardized ENEM score and corresponding 95 percent confidence intervals (*y*-axis) by days elapsed between the BFP disbursement date and the ENEM exam (*x*-axis). Test scores are standardized to have mean zero and standard deviation one among test-takers who received BFP.

Figure 4: ENEM score in each exam subject

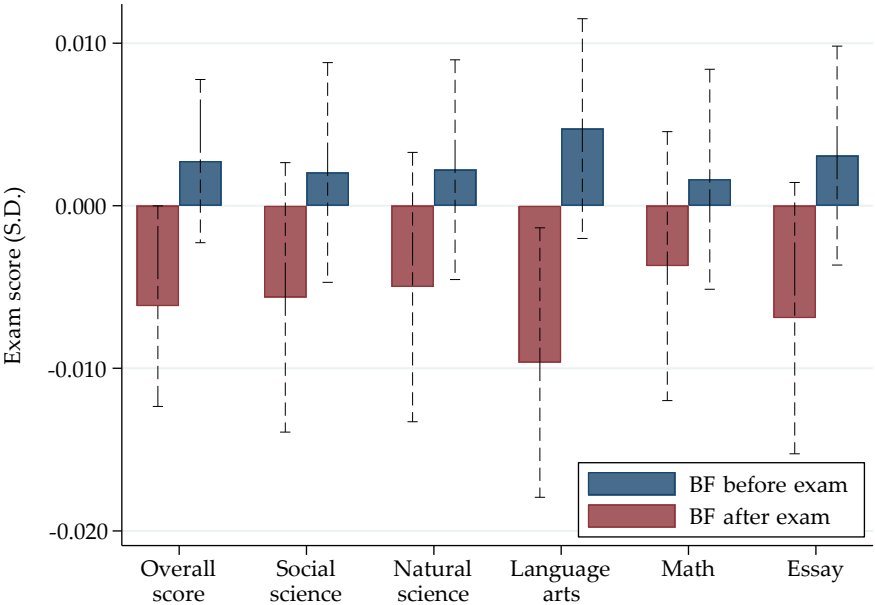

*Notes*: This figure shows the average standardized ENEM scores of students in households that received the BFP transfer before the exam (blue bars) to those who received it after (red bars). The overall score is computed as the average across the five subjects. Test scores are standardized to have mean zero and standard deviation one among test-takers who received BFP. Vertical dashed lines show 95 percent confidence intervals.

Table 1: Summary statistics on students in the sample

| | High school seniors (1) | In Cadastro Unico? Yes (2) | In Cadastro Unico? No (3) | In Bolsa Familia? Yes (4) | In Bolsa Familia? No (5) |
|---|---|---|---|---|---|
| **Panel A. Demographic characteristics and race** | | | | | |
| Age | 18.864 | 20.575 | 18.435 | 20.520 | 18.650 |
| Female | 0.588 | 0.632 | 0.577 | 0.652 | 0.580 |
| White | 0.465 | 0.324 | 0.500 | 0.276 | 0.489 |
| Black/Brown | 0.535 | 0.676 | 0.500 | 0.724 | 0.511 |
| **Panel B. Socioeconomic and household characteristics** | | | | | |
| Went to a private HS | 0.186 | 0.026 | 0.226 | 0.018 | 0.207 |
| Family size | 4.209 | 4.388 | 4.164 | 4.601 | 4.158 |
| Mom completed high school | 0.463 | 0.244 | 0.516 | 0.202 | 0.496 |
| Mom completed college | 0.162 | 0.033 | 0.193 | 0.020 | 0.179 |
| Family earns above 2x M.W. | 0.362 | 0.120 | 0.422 | 0.069 | 0.399 |
| Family earns above 5x M.W. | 0.043 | 0.001 | 0.053 | 0.000 | 0.048 |
| **Panel C. Geographic location** | | | | | |
| Lives in the North | 0.092 | 0.100 | 0.090 | 0.119 | 0.088 |
| Lives in the Northeast | 0.275 | 0.392 | 0.245 | 0.486 | 0.247 |
| Lives in the Southeast | 0.411 | 0.326 | 0.432 | 0.266 | 0.430 |
| Lives in the South | 0.139 | 0.097 | 0.149 | 0.061 | 0.149 |
| Lives in the Midwest | 0.084 | 0.085 | 0.083 | 0.068 | 0.086 |
| **Panel D. Exam attendance and test scores** | | | | | |
| Took at least one test | 0.855 | 0.782 | 0.873 | 0.784 | 0.864 |
| Took all tests | 0.832 | 0.753 | 0.852 | 0.757 | 0.842 |
| Overall score | 0.541 | 0.072 | 0.645 | -0.001 | 0.604 |
| Math | 0.610 | 0.082 | 0.727 | -0.001 | 0.681 |
| Natural science | 0.611 | 0.074 | 0.730 | -0.001 | 0.682 |
| Social science | 0.539 | 0.073 | 0.642 | -0.001 | 0.602 |
| Language arts | 0.553 | 0.082 | 0.657 | -0.001 | 0.617 |
| Essay | 0.392 | 0.046 | 0.468 | -0.001 | 0.437 |
| **Panel E. BFP benefits** | | | | | |
| Disbursed through bank deposit | 0.257 | 0.257 | . | 0.257 | . |
| Total BF benefits value (US$) | 94.379 | 94.379 | . | 94.379 | . |
| Number of test-takers | 1,616,321 | 324,404 | 1,291,917 | 184,895 | 1,431,426 |

*Notes*: This table shows the characteristics of high school seniors who registered for the ENEM exam (column 1), by household registration in the Registry of Social Programs (columns 2-3) and by BFP beneficiary status (columns 4-5). Panels A through D use ENEM administrative data, including students' socioeconomic questionnaire responses. Test scores are standardized to have mean zero and standard deviation one among BFP recipients (our estimating sample). Panel E uses administrative data from the Registry of Social Programs (Cadastro Unico) and BFP program records.

Table 2: Balance test of students receiving BFP before versus after the exam

|  | Receives Bolsa Familia | | |
|---|---|---|---|
|  | Before exam (1) | After exam (2) | Difference (3) |
| **Panel A. Student characteristics** | | | |
| Age | 20.50 | 20.54 | -0.036 |
| Female | 0.652 | 0.651 | 0.001 |
| White | 0.276 | 0.276 | 0.000 |
| Went to a private HS | 0.018 | 0.018 | -0.000 |
| **Panel B. Socioeconomic and household characteristics** | | | |
| Monthly income (US$) | 197.43 | 198.57 | -1.138 |
| Monthly p.c. income (US$) | 47.73 | 47.90 | -0.176 |
| Family size | 4.602 | 4.600 | 0.001 |
| 0-17 year olds in household | 1.714 | 1.716 | -0.002 |
| 0-2 year olds in household | 0.092 | 0.092 | 0.000 |
| Share of women in household | 0.589 | 0.590 | -0.001 |
| Mom completed high school | 0.202 | 0.202 | -0.000 |
| Mom completed college | 0.020 | 0.020 | 0.000 |
| **Panel C. Geographical location** | | | |
| Lives in the North | 0.119 | 0.121 | -0.002 |
| Lives in the Northeast | 0.486 | 0.485 | 0.001 |
| Lives in the Southeast | 0.266 | 0.265 | 0.002 |
| Lives in the South | 0.061 | 0.062 | -0.001 |
| Lives in the Midwest | 0.068 | 0.068 | -0.000 |
| **Panel D. BFP benefits** | | | |
| Disbursed through bank deposit | 0.258 | 0.254 | 0.004* |
| BF Transfer Size (US$) | 94.35 | 94.43 | -0.080 |
| **Panel E. Predicted score** | | | |
| Predicted mean score | -0.022 | -0.023 | 0.001 |
| Number of test-takers | 111,201 | 73,694 | 184,895 |

*Notes*: This table compares characteristics of students in households that received BFP before the exam (column 1) versus after the exam (column 2). Column 3 reports coefficients from student-level regressions of each characteristic on an indicator for receiving BFP before the exam. *, **, and *** denote significance at the 1%, 5%, and 10% levels, respectively.

Table 3: The effect of BFP payment timing on the probability of attending the ENEM

|  | Dependent variable: | | | |
|---|---|---|---|---|
|  | Attended both days (1) | Attended any day (2) | Attended day 1 (3) | Attended day 2 (4) |
| BF before exam | -0.001 | -0.002 | -0.002 | -0.001 |
|  | (0.002) | (0.002) | (0.002) | (0.002) |
| Mean after | 0.758 | 0.785 | 0.782 | 0.761 |
| $N$ | 184,895 | 184,895 | 184,895 | 184,895 |

*Notes*: This table reports estimates of $\beta$ from equation (1). The sample includes all final-year high school students who registered for the ENEM. The outcomes are indicator variables for attending both exam days (column 1), attending any exam day (column 2), attending the first day (column 3), and attending the second day (column 4). Heteroskedasticity-robust standard errors clustered at the student level in parentheses. *, **, and *** denote significance at the 10%, 5%, and 1% levels, respectively.

Table 4: The effect of BFP payment timing on test scores

| | Dependent variable: | | | | | |
|---|---|---|---|---|---|---|
| | Average score | Social science | Natural science | Lang. arts | Math | Essay |
| | (1) | (2) | (3) | (4) | (5) | (6) |
| BF before exam | 0.009** | 0.008 | 0.007 | 0.014*** | 0.005 | 0.010* |
| | (0.004) | (0.005) | (0.005) | (0.005) | (0.005) | (0.005) |
| Mean after | -0.006 | -0.006 | -0.005 | -0.010 | -0.004 | -0.007 |
| $N$ | 140,040 | 140,040 | 140,040 | 140,040 | 140,040 | 140,040 |

*Notes*: This table reports estimates of $\beta$ from equation (1). The sample includes final-year high school students who attended both exam days. The outcomes are the average standardized test score across all subjects (column 1) and the standardized test score in each subject (columns 2-6). Test scores are standardized to have mean zero and standard deviation one among BFP recipients (our estimating sample). Heteroskedasticity-robust standard errors clustered at the student level in parentheses. *, **, and *** denote significance at the 10%, 5%, and 1% levels, respectively.

Table 5: The effect of BFP payment timing on test scores by transfer size

| | Dependent variable: | | | | | |
|---|---|---|---|---|---|---|
| | Average score | Social science | Natural science | Lang. arts | Math | Essay |
| | (1) | (2) | (3) | (4) | (5) | (6) |
| **Panel A. Top transfer size tercile** | | | | | | |
| BF before exam | 0.020*** | 0.020** | 0.011 | 0.026** | 0.020** | 0.025** |
| | (0.007) | (0.010) | (0.010) | (0.010) | (0.010) | (0.010) |
| Mean after | -0.089 | -0.088 | -0.071 | -0.113 | -0.090 | -0.083 |
| $N$ | 40,521 | 40,521 | 40,521 | 40,521 | 40,521 | 40,521 |
| **Panel B. Middle transfer size tercile** | | | | | | |
| BF before exam | 0.000 | -0.005 | 0.000 | 0.002 | -0.001 | 0.006 |
| | (0.007) | (0.009) | (0.010) | (0.010) | (0.010) | (0.009) |
| Mean after | 0.019 | 0.023 | 0.012 | 0.024 | 0.019 | 0.014 |
| $N$ | 45,912 | 45,912 | 45,912 | 45,912 | 45,912 | 45,912 |
| **Panel C. Bottom transfer size tercile** | | | | | | |
| BF before exam | 0.008 | 0.010 | 0.011 | 0.016* | -0.000 | 0.002 |
| | (0.007) | (0.009) | (0.009) | (0.009) | (0.009) | (0.009) |
| Mean after | 0.035 | 0.033 | 0.030 | 0.039 | 0.042 | 0.033 |
| $N$ | 53,607 | 53,607 | 53,607 | 53,607 | 53,607 | 53,607 |

*Notes*: This table reports estimates of $\beta$ from equation (1) separately for three groups of students: those receiving transfers in the top size tercile (Panel A), middle size tercile (Panel B), and bottom size tercile (Panel C). The sample includes final-year high school students who attended both exam days. The outcomes are the average standardized test score across all subjects (column 1) and the standardized test score in each subject (columns 2-6). Test scores are standardized to have mean zero and standard deviation one among BFP recipients (our estimating sample). Heteroskedasticity-robust standard errors clustered at the student level in parentheses. *, **, and *** denote significance at the 10%, 5%, and 1% levels, respectively.

Table 6: The effect of BFP timing on college enrollment, college quality, and graduation

|  | Dependent variable: | | | | | |
|---|---|---|---|---|---|---|
|  | Enroll college (1) | Private college (2) | Public college (3) | College Quality (4) | 3-year persist (5) | 7-year grad (6) |
| **Panel A. All BFP test-takers** | | | | | | |
| BF before exam | 0.002 | 0.000 | 0.001 | 0.787 | 0.003 | 0.001 |
|  | (0.002) | (0.002) | (0.001) | (1.408) | (0.002) | (0.002) |
| Mean after | 0.181 | 0.134 | 0.050 | 372.876 | 0.232 | 0.158 |
| $N$ | 140,032 | 140,032 | 140,032 | 140,032 | 140,032 | 140,032 |
| **Panel B. Top transfer size tercile** | | | | | | |
| BF before exam | 0.006* | 0.005 | 0.001 | 5.410** | 0.008** | 0.008** |
|  | (0.004) | (0.003) | (0.002) | (2.626) | (0.004) | (0.003) |
| Mean after | 0.153 | 0.107 | 0.049 | 341.291 | 0.196 | 0.130 |
| $N$ | 40,515 | 40,515 | 40,515 | 40,516 | 40,515 | 40,515 |
| **Panel C. Middle transfer size tercile** | | | | | | |
| BF before exam | -0.004 | -0.005 | 0.001 | -2.028 | -0.003 | -0.002 |
|  | (0.004) | (0.003) | (0.002) | (2.630) | (0.004) | (0.004) |
| Mean after | 0.188 | 0.140 | 0.051 | 377.419 | 0.242 | 0.162 |
| $N$ | 45,911 | 45,911 | 45,911 | 45,910 | 45,911 | 45,911 |
| **Panel D. Bottom transfer size tercile** | | | | | | |
| BF before exam | 0.003 | 0.002 | 0.002 | -0.250 | 0.004 | 0.001 |
|  | (0.004) | (0.003) | (0.002) | (2.495) | (0.004) | (0.003) |
| Mean after | 0.197 | 0.148 | 0.051 | 392.840 | 0.252 | 0.176 |
| $N$ | 53,606 | 53,606 | 53,606 | 53,606 | 53,606 | 53,606 |

*Notes*: This table reports estimates of $\beta$ from equation (1) for four groups: all BFP test-takers (Panel A), and students receiving transfers in the top (Panel B), middle (Panel C), and bottom (Panel D) size terciles. The sample includes final-year high school students who attended both exam days. The outcomes are indicators for enrollment in any college (column 1), a private college (column 2), and a public college (column 3), a college quality index in USD (column 4), and indicators for persistence three years after enrollment (column 5) and graduation seven years after enrollment (column 6). See Appendix B.1 for variable definitions. Heteroskedasticity-robust standard errors clustered at the student level in parentheses. *, **, and *** denote significance at the 10%, 5%, and 1% levels, respectively.

Table 7: Testing college enrollment mechanisms among large-transfer recipients

|  | Dependent variable: | | | | | |
|---|---|---|---|---|---|---|
|  | Selective admit (1) | Scholarship requirements: | | | Enrolled with scholarship | |
|  |  | ProUni eligible (2) | ENEM over 450 (3) | Essay non-zero (4) | Partial tuition (5) | Full tuition (6) |
| BF before exam | -0.000 | 0.008 | 0.008 | 0.004** | 0.004 | 0.006 |
|  | (0.002) | (0.005) | (0.005) | (0.002) | (0.007) | (0.007) |
| Mean after | 0.044 | 0.468 | 0.468 | 0.954 | 0.085 | 0.070 |
| $N$ | 40,515 | 40,521 | 40,521 | 40,521 | 6,514 | 6,514 |

*Notes*: This table reports estimates of $\beta$ from equation (1) for students receiving transfers in the top size tercile. The sample includes final-year high school students who attended both exam days. Column headers show the outcomes: selective admission process participation (1), scholarship eligibility (2), meeting minimum ENEM score requirement (3), meeting minimum essay score requirement (4), and enrollment with partial (5) or full (6) tuition scholarship. Heteroskedasticity-robust standard errors clustered at the student level in parentheses. *, **, and *** denote significance at the 10%, 5%, and 1% levels, respectively.

Table 8: The effect of BFP payment timing on labor market outcomes

| | Dependent variable (measured 5–9 years after ENEM) | | | |
|---|---|---|---|---|
| | Formal employment (1) | Earnings (if employed) (2) | Earnings (unconditional) (3) | Hourly wage (if employed) (4) |
| **Panel A. All BFP test-takers** | | | | |
| BF before exam | -0.000 | -1.894 | -0.747 | 0.015 |
| | (0.003) | (2.203) | (0.905) | (0.023) |
| Mean after | 0.310 | 369.014 | 299.292 | 2.199 |
| $N$ | 140,040 | 47,683 | 140,040 | 47,669 |
| **Panel B. Top transfer size tercile** | | | | |
| BF before exam | 0.009* | -1.950 | 1.544 | -0.026 |
| | (0.005) | (3.888) | (1.541) | (0.040) |
| Mean after | 0.288 | 361.091 | 283.646 | 2.158 |
| $N$ | 40,521 | 13,135 | 40,521 | 13,132 |
| **Panel C. Middle transfer size tercile** | | | | |
| BF before exam | -0.000 | 0.286 | -0.653 | -0.005 |
| | (0.004) | (3.654) | (1.531) | (0.044) |
| Mean after | 0.312 | 365.198 | 299.311 | 2.223 |
| $N$ | 45,912 | 15,780 | 45,912 | 15,773 |
| **Panel D. Bottom transfer size tercile** | | | | |
| BF before exam | -0.007 | -3.292 | -2.503 | 0.061* |
| | (0.004) | (3.798) | (1.578) | (0.037) |
| Mean after | 0.324 | 377.513 | 311.065 | 2.208 |
| $N$ | 53,607 | 18,768 | 53,607 | 18,764 |

*Notes*: This table reports estimates of $\beta$ from equation (1) for four groups: all BFP test-takers (Panel A), and students receiving transfers in the top (Panel B), middle (Panel C), and bottom (Panel D) size terciles. The sample includes final-year high school students who attended both exam days. The outcomes, measured using RAIS administrative data during 2018–2022 (5–9 years after the exam), are: formal employment (indicator for any formal employment during the period), monthly earnings among the employed (average earnings in years with formal employment), monthly earnings including imputed informal sector earnings (where non-formal workers are assigned their state's mean informal sector earnings from PNAD Contínua), and hourly wages among the employed (monthly earnings divided by contracted monthly hours). See Appendix B.1 for variable definitions. Heteroskedasticity-robust standard errors clustered at the student level in parentheses. *, **, and *** denote significance at the 10%, 5%, and 1% levels, respectively.

# Appendix

## A  Appendix Figures and Tables

Figure A1: Distribution of NIS last digits among students in 2013

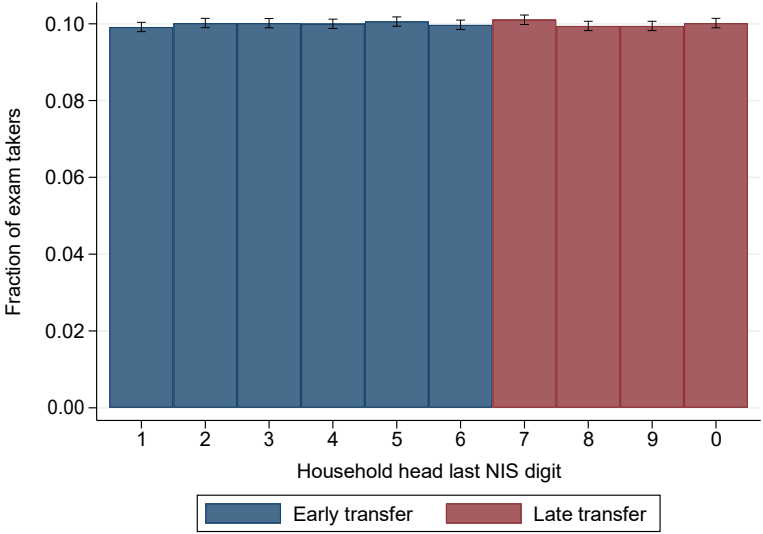

*Notes*: This figure shows the distribution of students by the last digit of their household's NIS in 2013. Blue bars represent households that received their BFP transfer before the ENEM exam, while red bars represent households that received it after.

Figure A2: Share of households withdrawing BFP before the ENEM by NIS last digit

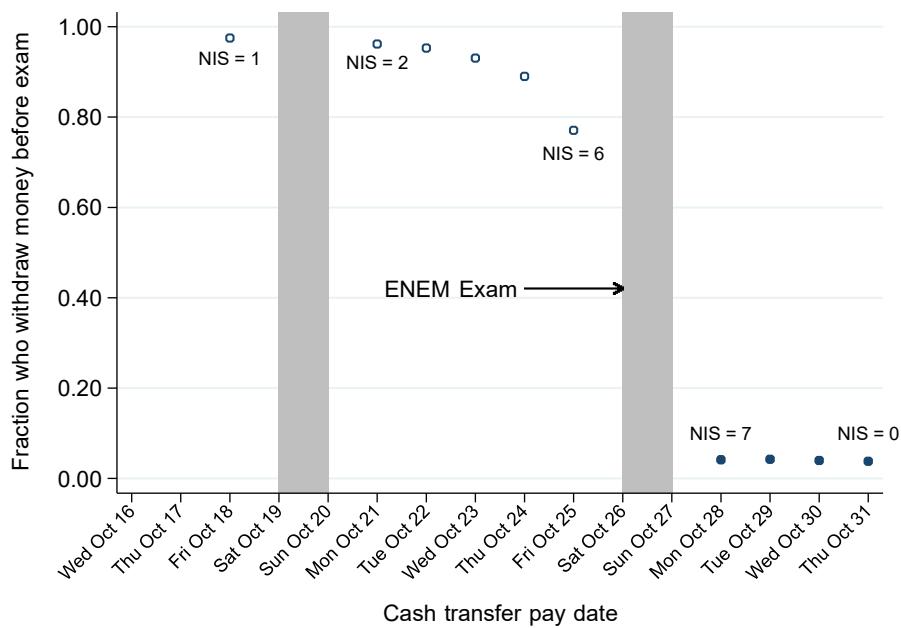

*Notes*: This figure shows the fraction of households that withdrew their October 2013 BFP payment by Friday, October 25 (the day before the ENEM exam), separately by NIS last digit. Payment disbursement began on Monday, October 21 for NIS ending in 1, proceeding sequentially on subsequent business days. Gray shaded areas indicate weekends.

Figure A3: ENEM attendance rates by NIS last digit

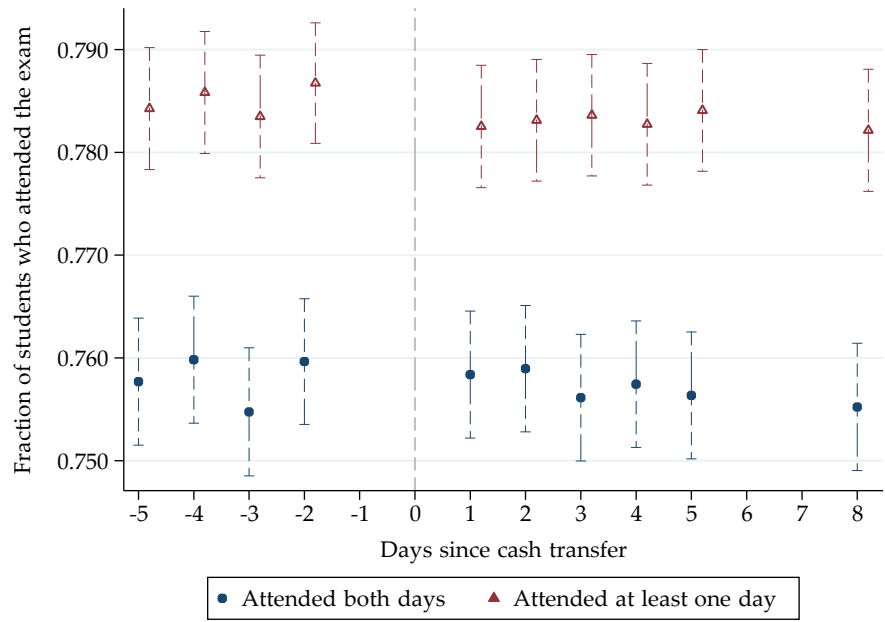

*Notes*: This figure shows the proportion of registered students who attended both exam days (blue circles) or at least one exam day (red triangles) by days between BFP disbursement and the exam (*x*-axis). Positive *x*-axis values indicate students whose households received BFP before the ENEM exam, while negative values indicate those who received it after.

Figure A4: Effect of BFP payment timing on performance by question position

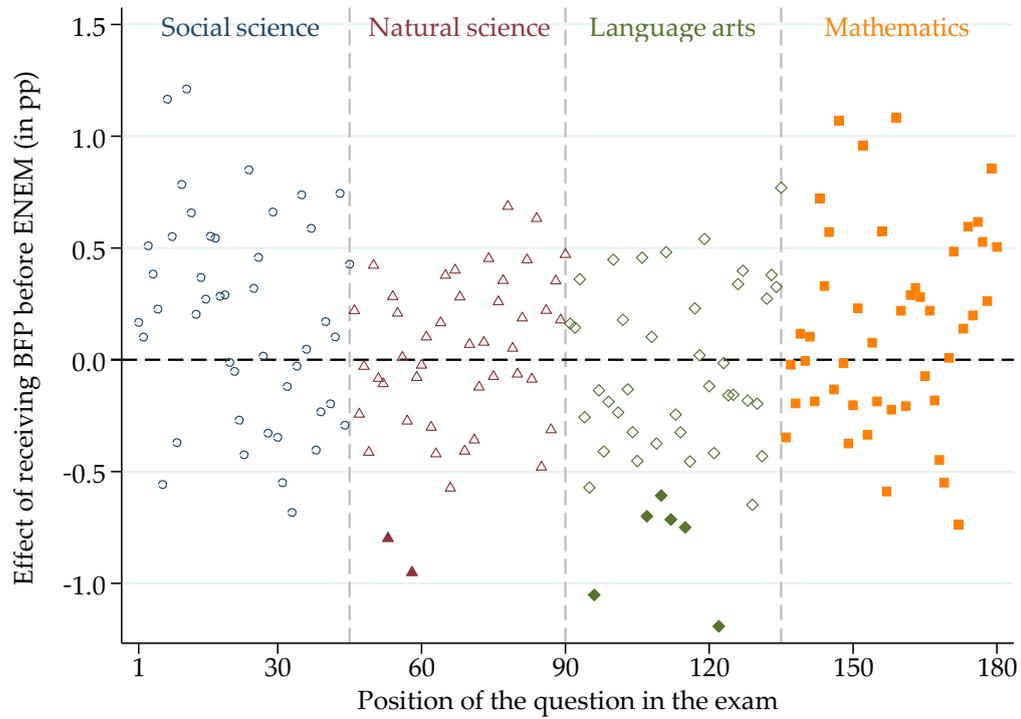

*Notes*: This figure shows the estimated effect (in percentage points) of receiving BFP before the ENEM exam on the probability of answering questions correctly. The *x*-axis shows question position, and the *y*-axis shows the estimated effect. Colors/markers indicate subject areas: Social Science (blue circles), Natural Science (red triangles), Language Arts (green diamonds), and Mathematics (orange squares). Dashed vertical lines separate subject sections. The shaded orange region highlights the Mathematics section, where effects are most pronounced.

Figure A5: Distribution of ENEM scores and public university admission cutoffs

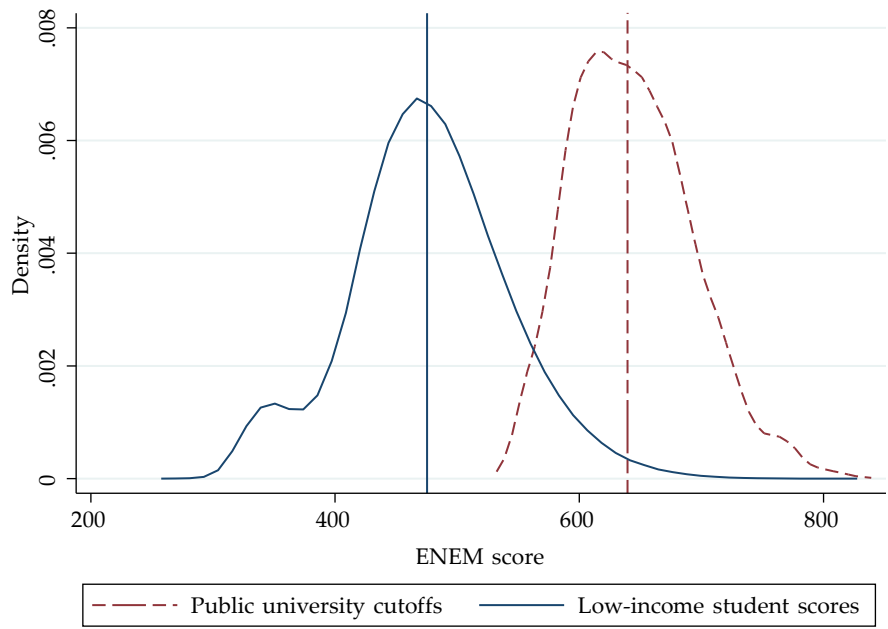

*Notes*: This figure shows the distribution of 2017 admission cutoff scores across public university programs (red dashed line) and the distribution of 2017 ENEM scores among students with family income below the minimum wage (solid blue line). Admission cutoffs come from SISU, the centralized system used by the Ministry of Education to allocate students to public universities based on ENEM scores. Vertical dashed and solid lines show median admission cutoff and ENEM scores, respectively.

Table A1: Average student characteristics by last NIS digit

| | Last NIS digit: | | | | | | | | | | | |
|---|---|---|---|---|---|---|---|---|---|---|---|---|
| | 1 | 2 | 3 | 4 | 5 | 6 | 7 | 8 | 9 | 0 | F | P-val. |
| **Panel A. Student characteristics** | | | | | | | | | | | | |
| Age | 20.45 | 20.64 | 20.53 | 20.42 | 20.54 | 20.46 | 20.51 | 20.55 | 20.56 | 20.55 | 2.05 | 0.030 |
| Female | 0.651 | 0.657 | 0.648 | 0.650 | 0.653 | 0.653 | 0.651 | 0.649 | 0.650 | 0.654 | 0.61 | 0.794 |
| White | 0.282 | 0.278 | 0.276 | 0.275 | 0.269 | 0.276 | 0.272 | 0.278 | 0.277 | 0.276 | 1.08 | 0.373 |
| Went to a private HS | 0.018 | 0.020 | 0.018 | 0.018 | 0.019 | 0.018 | 0.020 | 0.018 | 0.017 | 0.019 | 0.84 | 0.583 |
| **Panel B. SES and household characteristics** | | | | | | | | | | | | |
| Monthly income (US$) | 198.06 | 197.01 | 197.73 | 198.14 | 196.05 | 197.57 | 197.66 | 199.89 | 198.54 | 198.19 | 0.53 | 0.856 |
| Monthly p.c. income (US$) | 47.78 | 47.72 | 47.73 | 47.82 | 47.63 | 47.68 | 47.84 | 48.20 | 48.01 | 47.57 | 0.35 | 0.957 |
| Family size | 4.594 | 4.596 | 4.605 | 4.606 | 4.603 | 4.608 | 4.596 | 4.590 | 4.595 | 4.621 | 0.56 | 0.828 |
| 0-17 year olds in household | 1.714 | 1.717 | 1.708 | 1.717 | 1.714 | 1.712 | 1.701 | 1.717 | 1.711 | 1.737 | 1.12 | 0.344 |
| 0-2 year olds in household | 0.092 | 0.093 | 0.091 | 0.092 | 0.093 | 0.094 | 0.092 | 0.089 | 0.094 | 0.093 | 0.35 | 0.959 |
| Share of women in household | 0.589 | 0.592 | 0.587 | 0.593 | 0.589 | 0.588 | 0.592 | 0.587 | 0.588 | 0.592 | 1.80 | 0.062 |
| Mom completed high school | 0.196 | 0.202 | 0.206 | 0.205 | 0.202 | 0.200 | 0.200 | 0.201 | 0.203 | 0.204 | 1.01 | 0.433 |
| Mom completed college | 0.021 | 0.020 | 0.020 | 0.022 | 0.021 | 0.019 | 0.020 | 0.020 | 0.019 | 0.021 | 0.73 | 0.685 |
| **Panel C. Bolsa Fanilia Benefits** | | | | | | | | | | | | |
| Disbursed through bank deposit | 0.257 | 0.259 | 0.255 | 0.257 | 0.260 | 0.260 | 0.260 | 0.251 | 0.251 | 0.255 | 1.25 | 0.256 |
| BF Transfer Size (US$) | 94.30 | 94.33 | 94.26 | 94.64 | 94.72 | 93.82 | 93.96 | 94.17 | 94.58 | 95.01 | 0.86 | 0.559 |
| **Panel D. Geographical location** | | | | | | | | | | | | |
| Lives in the North | 0.123 | 0.116 | 0.121 | 0.117 | 0.117 | 0.118 | 0.117 | 0.119 | 0.122 | 0.124 | 1.39 | 0.186 |
| Lives in the Northeast | 0.482 | 0.492 | 0.484 | 0.489 | 0.494 | 0.478 | 0.490 | 0.484 | 0.484 | 0.481 | 1.92 | 0.044 |
| Lives in the Southeast | 0.265 | 0.263 | 0.268 | 0.268 | 0.260 | 0.274 | 0.261 | 0.268 | 0.265 | 0.267 | 1.67 | 0.091 |
| Lives in the South | 0.063 | 0.063 | 0.058 | 0.060 | 0.062 | 0.060 | 0.065 | 0.063 | 0.059 | 0.060 | 1.57 | 0.118 |
| Lives in the Midwest | 0.067 | 0.066 | 0.070 | 0.066 | 0.067 | 0.070 | 0.067 | 0.067 | 0.069 | 0.068 | 0.67 | 0.735 |
| **Panel E. Predicted Score** | | | | | | | | | | | | |
| Predicted mean score | -0.021 | -0.024 | -0.022 | -0.019 | -0.024 | -0.020 | -0.023 | -0.021 | -0.024 | -0.022 | 0.66 | 0.742 |
| Number of test-takers | 18,488 | 18,522 | 18,621 | 18,615 | 18,568 | 18,387 | 18,696 | 18,247 | 18,328 | 18,423 | | |

*Notes*: This table compares the average characteristics of students according to the last digit of the NIS number associated with the BFP beneficiary in a student's household. The last NIS digit fully determines the date in which a student's family receives the cash transfer. Each column reports the average characteristic of students associated with the digit denoted in the column header. For each NIS digit $d$, the F-statistic and associated p-value are obtained from regressing an indicator variable that takes the value of one if the BFP beneficiary in student $i$'s household has a last NIS digit $d$ on the characteristics shown in panels A-C.

Table A2: The effect of BFP payment timing on test scores with student controls

|  | Dependent variable: | | | | | |
|---|---|---|---|---|---|---|
|  | Average score | Social science | Natural science | Lang. arts | Math | Essay |
|  | (1) | (2) | (3) | (4) | (5) | (6) |
| BF before exam | 0.007* | 0.007 | 0.006 | 0.013** | 0.004 | 0.008 |
|  | (0.004) | (0.005) | (0.005) | (0.005) | (0.005) | (0.005) |
| Mean after | -0.006 | -0.006 | -0.005 | -0.010 | -0.004 | -0.007 |
| $N$ | 140,040 | 140,040 | 140,040 | 140,040 | 140,040 | 140,040 |
| Controls? | Yes | Yes | Yes | Yes | Yes | Yes |

*Notes*: This table reports estimates of $\beta$ from equation (1) controlling for student characteristics (age, gender, race, income, and parental education). The sample includes final-year high school students who attended both exam days. The outcomes are the average standardized test score across all subjects (column 1) and standardized test scores in each subject (columns 2-6). Test scores are standardized to have mean zero and standard deviation one among BFP recipients (our estimating sample). See Appendix B.1 for variable definitions. Heteroskedasticity-robust standard errors clustered at the student level in parentheses. *, **, and *** denote significance at the 10%, 5%, and 1% levels, respectively.

Table A3: Effect of BFP payment timing on essay score components

| | | Components of the essay score: | | | | |
|---|---|---|---|---|---|---|
| | Overall Score (1) | Language Proficiency (2) | Theme Development (3) | Argument Construction (4) | Coherence and Cohesion (5) | Intervention Proposal (6) |
| **Panel A. All test-takers** | | | | | | |
| BF before exam | 0.010* | 0.012** | 0.007 | 0.007 | 0.011** | 0.005 |
| | (0.005) | (0.005) | (0.005) | (0.005) | (0.005) | (0.005) |
| Mean after | -0.007 | -0.008 | -0.005 | -0.005 | -0.007 | -0.004 |
| N | 140,040 | 140,040 | 140,040 | 140,040 | 140,040 | 140,040 |
| **Panel B. Top transfer size tercile** | | | | | | |
| BF before exam | 0.025** | 0.025** | 0.027*** | 0.019* | 0.026** | 0.010 |
| | (0.010) | (0.010) | (0.010) | (0.010) | (0.010) | (0.010) |
| Mean after | -0.083 | -0.081 | -0.076 | -0.077 | -0.077 | -0.042 |
| N | 40,521 | 40,521 | 40,521 | 40,521 | 40,521 | 40,521 |
| **Panel C. Middle transfer size tercile** | | | | | | |
| BF before exam | 0.006 | 0.007 | 0.005 | 0.007 | 0.011 | -0.003 |
| | (0.009) | (0.009) | (0.009) | (0.009) | (0.009) | (0.009) |
| Mean after | 0.014 | 0.014 | 0.010 | 0.013 | 0.009 | 0.012 |
| N | 45,912 | 45,912 | 45,912 | 45,912 | 45,912 | 45,912 |
| **Panel D. Bottom transfer size tercile** | | | | | | |
| BF before exam | 0.002 | 0.008 | -0.006 | -0.003 | -0.001 | 0.009 |
| | (0.009) | (0.009) | (0.009) | (0.009) | (0.009) | (0.009) |
| Mean after | 0.033 | 0.028 | 0.036 | 0.034 | 0.031 | 0.012 |
| N | 53,607 | 53,607 | 53,607 | 53,607 | 53,607 | 53,607 |

*Notes*: This table reports estimates of $\beta$ from equation (1) for four groups: all test-takers (Panel A) and students receiving transfers in the top (Panel B), middle (Panel C), and bottom (Panel D) size terciles. The outcomes include the overall essay score (column 1) and scores on five specific competencies. *Language Proficiency* assesses mastery of formal written Portuguese, including grammar, spelling, punctuation, and syntactic accuracy. *Theme Development* evaluates comprehension of the essay prompt, adherence to the assigned topic, and integration of relevant disciplinary knowledge. *Argument Construction* measures the ability to develop a clear thesis, present well-structured arguments, and use supporting evidence effectively. *Coherence and Cohesion* examines the logical sequencing of ideas, textual unity, and appropriate use of linguistic mechanisms for connection and transition. *Intervention Proposal* assesses the development of a feasible and well-structured social intervention proposal, ensuring it aligns with the discussion and respects human rights. All scores are standardized to have a mean of zero and a standard deviation of one among BFP recipients (our estimating sample). Heteroskedasticity-robust standard errors clustered at the student level are shown in parentheses. *, **, and *** denote significance at the 10%, 5%, and 1% levels, respectively.

Table A4: Placebo test: Effect of last NIS digit on test scores among Social Registry enrollees not in BFP

|  | Dependent variable: | | | | | |
|---|---|---|---|---|---|---|
|  | Average score | Social science | Natural science | Lang. arts | Math | Essay |
|  | (1) | (2) | (3) | (4) | (5) | (6) |
| BF before exam | 0.007 | 0.004 | 0.006 | 0.006 | 0.010 | 0.008 |
|  | (0.005) | (0.007) | (0.007) | (0.007) | (0.007) | (0.006) |
| Mean after | 0.167 | 0.174 | 0.174 | 0.194 | 0.189 | 0.106 |
| N | 100,334 | 100,334 | 100,334 | 100,334 | 100,334 | 100,334 |

*Notes*: This table reports estimates of $\beta$ from a placebo version of equation (1). The sample includes students with a household member enrolled in the Registry of Social Programs (Cadastro Único) who was not a BFP beneficiary. We construct a placebo treatment variable that equals one if the household member's NIS last digit would have made them eligible for BFP receipt before the exam had they been enrolled in BFP. The outcomes are the average standardized test score across all subjects (column 1) and standardized test scores in each subject (columns 2-6). Test scores are standardized to have mean zero and standard deviation one among BFP recipients (our estimating sample). See Appendix B.1 for variable definitions. Heteroskedasticity-robust standard errors clustered at the student level are shown in parentheses. *, **, and *** denote significance at the 10%, 5%, and 1% levels, respectively.

Table A5: Household characteristics by transfer size tercile

|  | Transfer size tercile | | |
| --- | --- | --- | --- |
|  | Tercile 1 (Bottom) | Tercile 2 (Middle) | Tercile 3 (Top) |
| Transfer size (USD) | 51 | 92 | 175 |
| Monthly income (USD) | 234 | 179 | 105 |
| Monthly income per capita (USD) | 64 | 40 | 18 |
| BFP share of income | 0.26 | 0.45 | 0.72 |
| Household size (# members) | 3.49 | 4.12 | 5.48 |

*Notes*: This table reports average household characteristics by transfer size tercile. Transfer size refers to the monthly BFP benefit amount. Monthly income and income per capita are measured in USD. BFP share of income is the ratio of monthly BFP benefits to total household monthly income.

Table A6: Effect of BFP payment timing on test scores by transfer tercile controlling for student characteristics

| | Dependent variable: | | | | | |
|---|---|---|---|---|---|---|
| | Average score (1) | Social science (2) | Natural science (3) | Lang. arts (4) | Math (5) | Essay (6) |
| **Panel A. Top transfer size tercile** | | | | | | |
| BF before exam | 0.018** | 0.018* | 0.009 | 0.023** | 0.017* | 0.021** |
| | (0.007) | (0.010) | (0.010) | (0.010) | (0.009) | (0.010) |
| Mean after | -0.089 | -0.088 | -0.071 | -0.113 | -0.090 | -0.083 |
| $N$ | 40,521 | 40,521 | 40,521 | 40,521 | 40,521 | 40,521 |
| **Panel B. Middle transfer size tercile** | | | | | | |
| BF before exam | -0.002 | -0.007 | -0.002 | 0.000 | -0.003 | 0.003 |
| | (0.007) | (0.009) | (0.009) | (0.009) | (0.009) | (0.009) |
| Mean after | 0.019 | 0.023 | 0.012 | 0.024 | 0.019 | 0.014 |
| $N$ | 45,912 | 45,912 | 45,912 | 45,912 | 45,912 | 45,912 |
| **Panel C. Bottom transfer size tercile** | | | | | | |
| BF before exam | 0.008 | 0.010 | 0.011 | 0.017** | 0.000 | 0.002 |
| | (0.006) | (0.009) | (0.009) | (0.009) | (0.008) | (0.008) |
| Mean after | 0.035 | 0.033 | 0.030 | 0.039 | 0.042 | 0.033 |
| $N$ | 53,607 | 53,607 | 53,607 | 53,607 | 53,607 | 53,607 |
| Controls? | Yes | Yes | Yes | Yes | Yes | Yes |

*Notes*: This table reports estimates of $\beta$ from equation (1) controlling for students' socioeconomic and demographic characteristics, separately for three groups: students receiving transfers in the top size tercile (Panel A), middle size tercile (Panel B), and bottom size tercile (Panel C). The sample includes final-year high school students who attended both exam days. The outcomes are the average standardized test score across all subjects (column 1) and standardized test scores in each subject (columns 2-6). Test scores are standardized to have mean zero and standard deviation one among BFP recipients (our estimating sample). See Appendix B.1 for variable definitions. Heteroskedasticity-robust standard errors clustered at the student level are shown in parentheses. *, **, and *** denote significance at the 10%, 5%, and 1% levels, respectively.

Table A7: Effect of BFP payment timing on test scores by household composition

| | Heterogeneity dimension: | | | |
|---|---|---|---|---|
| | Above-median female members (1) | Above-median members age 0-2 (2) | Above-median members age 0-5 (3) | Above-median members age 18-24 (4) |
| BF before exam | 0.010* | 0.008* | 0.009* | 0.009** |
| | (0.005) | (0.004) | (0.005) | (0.004) |
| Before × het. | -0.002 | 0.010 | 0.001 | 0.001 |
| | (0.008) | (0.014) | (0.010) | (0.016) |
| $N$ | 140,040 | 140,040 | 140,040 | 140,040 |

*Notes*: This table reports estimates from augmented versions of equation (1) that include indicators for household composition and their interactions with receiving BFP before the exam. The outcome variable is the standardized ENEM score. "BF before exam" shows the main effect of receiving transfers before the exam, while "Before × het." shows the interaction effect with indicators for: above-median number of female household members (column 1), above-median number of household members aged 0-2 (column 2), above-median number of household members aged 0-5 (column 3), and above-median number of household members aged 18-24 (column 4). Standardized test scores are standardized to have mean zero and standard deviation one among BFP recipients (our estimating sample). Heteroskedasticity-robust standard errors clustered at the student level are shown in parentheses. *, **, and *** denote significance at the 10%, 5%, and 1% levels, respectively.

Table A8: Effect of BFP payment timing on test scores by student characteristics

|  | Heterogeneity dimension: | | | | |
|---|---|---|---|---|---|
|  | Works > 20 hours/week (1) | College-educated mother (2) | Female student (3) | White student (4) | Same muni as exam (5) |
| BF before exam | 0.010** | 0.008* | 0.008 | 0.007 | 0.006 |
|  | (0.004) | (0.004) | (0.007) | (0.005) | (0.008) |
| Before × het. | -0.006 | 0.057* | 0.001 | 0.005 | 0.004 |
|  | (0.013) | (0.031) | (0.009) | (0.009) | (0.010) |
| $N$ | 140,040 | 140,040 | 140,040 | 140,040 | 140,040 |

*Notes*: This table reports estimates from augmented versions of equation (1) that include indicators for student characteristics and their interactions with receiving BFP before the exam. The outcome variable is the standardized ENEM score. "BFP before exam" shows the main effect of receiving transfers before the exam, while "Before × het." shows the interaction effect with indicators for: working more than 20 hours per week (column 1), having a college-educated mother (column 2), being female (column 3), being white (column 4), and taking the exam in municipality of residence (column 5). Test scores are standardized to have mean zero and standard deviation one among BFP recipients (our estimating sample). Heteroskedasticity-robust standard errors clustered at the student level are shown in parentheses. *, **, and *** denote significance at the 10%, 5%, and 1% levels, respectively.

Table A9: Effect of BFP payment timing on performance by question position

|  | Dependent variable: Indicator for correct answer | | |
|---|---|---|---|
|  | Last ten questions (any day) (1) | Last ten questions (day 1) (2) | Last ten questions (day 2) (3) |
| BF before exam | 0.000 | 0.000 | 0.000 |
|  | (0.000) | (0.000) | (0.000) |
| Before × last ten | 0.002** | 0.001 | 0.003** |
|  | (0.001) | (0.001) | (0.001) |
| Constant | 0.240*** | 0.238*** | 0.239*** |
|  | (0.000) | (0.000) | (0.000) |
| $N$ | 7,494,350 | 7,494,350 | 7,494,350 |

*Notes*: This table reports estimates from equation (2). The dependent variable is an indicator for whether the student correctly answered the question. "BF before exam" shows the main effect of receiving transfers before the exam, while "Before × last ten" shows the interaction effect with indicators for questions appearing in: the last ten questions of either test day (column 1), the last ten questions of day 1 (column 2), and the last ten questions of day 2 (column 3). The sample includes all question-level observations for students who received transfers in the top tercile of the transfer size distribution. Heteroskedasticity-robust standard errors clustered at the student level are shown in parentheses. *, **, and *** denote significance at the 10%, 5%, and 1% levels, respectively.

Table A10: Effect of BFP payment timing on performance by question difficulty

| | Share of correct responses in question difficulty percentile range: | | | | | |
|---|---|---|---|---|---|---|
| | 1st–10th | 10th–25th | 25th–50th | 50th–75th | 75th–90th | 90th–100th |
| | (1) | (2) | (3) | (4) | (5) | (6) |
| BF before exam | -0.0016* | -0.0014* | 0.0005 | 0.0008 | 0.0019 | 0.0039** |
| | (0.0008) | (0.0008) | (0.0011) | (0.0015) | (0.0014) | (0.0017) |
| Mean after | 0.131 | 0.160 | 0.342 | 0.453 | 0.347 | 0.463 |
| $N$ | 40,521 | 40,521 | 40,521 | 40,521 | 40,521 | 40,521 |

*Notes*: This table reports estimates of $\beta$ from equation (1) examining the effect of receiving BFP before the ENEM exam on performance across questions of different difficulty levels. The outcomes are the share of correct responses within each question difficulty percentile range, shown in the column headers. The sample includes students who received transfers in the top tercile of the transfer size distribution. Question difficulty is defined by the share of correct responses among late recipients. Heteroskedasticity-robust standard errors clustered at the student level are shown in parentheses. *, **, and *** denote significance at the 10%, 5%, and 1% levels, respectively.

Table A11: Effect of BFP payment timing on earnings-based college quality

|  | Dependent variable: Earnings-based college quality index | |
|---|---|---|
|  | Imputing earnings of students outside formal sector (1) | Students in formal sector only (2) |
| **Panel A. All test-takers** | | |
| BF before exam | 0.503 | 0.787 |
|  | (0.759) | (1.498) |
| Mean after | 294.618 | 372.876 |
| $N$ | 140,032 | 140,032 |
| **Panel B. Top transfer size tercile** | | |
| BF before exam | 4.021*** | 5.410** |
|  | (1.348) | (2.626) |
| Mean after | 275.178 | 341.291 |
| $N$ | 40,516 | 40,516 |
| **Panel C. Middle transfer size tercile** | | |
| BF before exam | -0.097 | -2.028 |
|  | (1.324) | (2.630) |
| Mean after | 296.092 | 377.419 |
| $N$ | 45,910 | 45,910 |
| **Panel D. Bottom transfer size tercile** | | |
| BF before exam | -1.597 | -0.250 |
|  | (1.256) | (2.495) |
| Mean after | 308.021 | 392.840 |
| $N$ | 53,606 | 53,606 |

*Notes*: This table reports estimates of $\beta$ from equation (1) for four groups: all test-takers (Panel A) and students receiving transfers in the top (Panel B), middle (Panel C), and bottom (Panel D) size terciles. The outcome is an earnings-based index of college quality, measured in USD. For each college program, we calculate mean earnings of graduates aged 25-35 in the RAIS formal-sector employment data. Column 1 then imputes earnings for students who are not observed in the formal sector using state-specific mean earnings of workers with similar education levels from the PNAD Contínua household survey. Column 2 restricts the sample to graduates working in the formal sector only. See Appendix B.1 for variable definitions. Heteroskedasticity-robust standard errors clustered at the student level are shown in parentheses. *, **, and *** denote significance at the 10%, 5%, and 1% levels, respectively.

Table A12: Effect of BFP payment timing on probability of scoring above different thresholds

|  | Dependent variable: Indicator for score above threshold | | | | | | | |
|---|---|---|---|---|---|---|---|---|
|  | 300 | 350 | 400 | 450 | 500 | 550 | 600 | 650 |
|  | (1) | (2) | (3) | (4) | (5) | (6) | (7) | (8) |
| **Panel A. All test-takers** | | | | | | | | |
| BF before exam | 0.001 | 0.002* | 0.003 | 0.006** | 0.004* | -0.000 | 0.000 | -0.000 |
|  | (0.000) | (0.001) | (0.002) | (0.003) | (0.002) | (0.001) | (0.001) | (0.000) |
| Mean after | 0.995 | 0.956 | 0.817 | 0.511 | 0.228 | 0.075 | 0.018 | 0.004 |
| $N$ | 140,040 | 140,040 | 140,040 | 140,040 | 140,040 | 140,040 | 140,040 | 140,040 |
| **Panel B. Top transfer size tercile** | | | | | | | | |
| BF before exam | 0.002*** | 0.004* | 0.010** | 0.008 | 0.009** | 0.002 | 0.000 | -0.000 |
|  | (0.001) | (0.002) | (0.004) | (0.005) | (0.004) | (0.002) | (0.001) | (0.001) |
| Mean after | 0.994 | 0.947 | 0.787 | 0.468 | 0.193 | 0.060 | 0.014 | 0.003 |
| $N$ | 40,521 | 40,521 | 40,521 | 40,521 | 40,521 | 40,521 | 40,521 | 40,521 |
| **Panel C. Middle transfer size tercile** | | | | | | | | |
| BF before exam | 0.001 | 0.002 | 0.001 | 0.006 | -0.001 | -0.003 | 0.001 | -0.001 |
|  | (0.001) | (0.002) | (0.004) | (0.005) | (0.004) | (0.003) | (0.001) | (0.001) |
| Mean after | 0.996 | 0.960 | 0.827 | 0.524 | 0.237 | 0.079 | 0.019 | 0.004 |
| $N$ | 45,912 | 45,912 | 45,912 | 45,912 | 45,912 | 45,912 | 45,912 | 45,912 |
| **Panel D. Bottom transfer size tercile** | | | | | | | | |
| BF before exam | -0.001 | 0.001 | -0.000 | 0.004 | 0.005 | 0.001 | -0.000 | 0.000 |
|  | (0.001) | (0.002) | (0.003) | (0.004) | (0.004) | (0.002) | (0.001) | (0.001) |
| Mean after | 0.996 | 0.961 | 0.831 | 0.533 | 0.245 | 0.082 | 0.022 | 0.004 |
| $N$ | 53,607 | 53,607 | 53,607 | 53,607 | 53,607 | 53,607 | 53,607 | 53,607 |

*Notes*: This table reports estimates of $\beta$ from equation (1) for four groups: all test-takers (Panel A) and students receiving transfers in the top (Panel B), middle (Panel C), and bottom (Panel D) size terciles. The sample includes final-year high school students who attended both exam days. The outcomes are indicators for whether a student's average ENEM score exceeded various thresholds, ranging from 300 to 650 points. Scores are on the original ENEM scale, which has a mean of 500 and standard deviation of 100 among all 2009 high school seniors who took the exam. Heteroskedasticity-robust standard errors clustered at the student level are shown in parentheses. *, **, and *** denote significance at the 10%, 5%, and 1% levels, respectively.

# B  Empirical Appendix

## B.1  Definition of Key Variables

*Test score.* The Brazilian Testing Agency grades the ENEM exam based on the three-parameter item response theory (IRT). According to IRT, the probability that an individual $i$ with ability $\theta_i$ correctly answers question $j$ is:

$$\Pr(C_{ij} = 1|\theta_i) = p_j(\theta_i) = c_j + \frac{1-c_j}{1+e^{-a_j(\theta_i-b_i)}}, \tag{B1}$$

where $C_{ij}$ indicates a correct response by student $i$ to question $j$, and parameters $a_j$, $b_j$, and $c_j$ capture question discrimination, difficulty, and pseudo-guessing probability, respectively.[14] ENEM scores, as reported by the testing agency, have mean 500 and standard deviation 100. We normalize test scores to have mean zero and standard deviation one in our sample.

*College enrollment.* We define college enrollment as an indicator for appearing in the higher-education census one year after taking the ENEM.

*College quality.* We construct an earnings-based index of college quality. define earnings-based indices of college quality. To do this, we group all college-educated workers in the employee-employer matched data for the years 2016-2022 based on the university they attended and compute the average predicted earnings at age 30 of the graduates from each university. For college graduates that are never observed in the RAIS, we assign the average monthly earnings in the informal sector in their state of residence. Students that do not attend a college are likewise assigned the informal sector earnings in their state. We calculate informal earnings in each state using the PNAD Contínua, a quarterly nationally-representative household survey.

*First-year persistence.* We define first-year persistence as an indicator that equals one if a student remains enrolled in college at the end of their freshmen year and zero otherwise. We create analogous measures for persistence in further years.

*Likelihood of graduating.* We define an indicator for graduating one to seven years after taking the ENEM. Most students who ever graduate do so within the first seven years.

*College financial assistance.* We define indicator variables that equal one for students who received different forms of financial assistance when enrolling in college in the year following the ENEM exam: (i) full or partial *ProUni* scholarships, which fund college-related expenses for low-income students attending private colleges; (ii) university-funded scholarships; (iii) government-funded student loans; (iv) university-funded student loans.

---

[14]A question's discrimination refers to its ability to differentiate between low- and high-ability individuals; difficulty represents the ability level at which $p_j(\theta_i)$ has maximum slope; and pseudo-guess indicates the probability that a student with extremely low ability correctly answers the question.

*Formal employment.* We define formal employment as an indicator for appearing in the employee-employer matched dataset during 2018–2022. This variable is defined for all students. If an individual has multiple jobs, we use the data from the job with the highest number of hours. We use the job monthly earnings as a tiebreaker.

*Monthly earnings.* This variable represents the average salary of a worker across all months in a given year. To report this variable, firms calculate the worker's total earnings for the year and divide them by the number of months the firm employed the worker. For workers appearing in multiple years in the RAIS, we calculate the inflation-adjusted average monthly earnings across all years in the 2018–2022 period. We adjust earnings for inflation using the consumer price index and express them in 2023 US Dollars. We define two versions of this measure: (1) formal-sector earnings for test-takers who appear in the RAIS, and (2) an unconditional version where we impute earnings for test-takers never observed in the RAIS using informal-sector earnings from *PNAD Contínua*, a quarterly, nationally-representative household survey. For the unconditional measure, we impute using average monthly earnings in the test-taker's state. Following the Brazilian Statistical Agency definition, we classify PNAD respondents as informal sector workers if they have unregistered employment relationships (without a signed work card), are self-employed or employers without a CNPJ (National Registry of Legal Entities), work as unregistered domestic workers, or engage in unpaid family work.

*Hourly wage.* We calculate the hourly rate of each worker as the ratio between a worker's inflation-adjusted monthly earnings and the hours worked per month.[15] If a worker appears in multiple years in the RAIS, we calculate the average hourly wage across all years.

---

[15]Firms do not record the number of hours individuals actually work each week. Instead, the data on hours indicates the number of hours per week that the worker is expected to work based on her contract.

# C  Conceptual framework

We present a simple framework linking monetary transfers to exam performance through cognitive functions. The framework builds on research showing that poverty can directly impair cognitive functioning (Mani et al., 2013; Schilbach et al., 2016) and that cognitive ability is a key determinant of test performance (Duckworth et al., 2011; Segal, 2012). Appendix Figure B1 illustrates these relationships.

The first component links economic resources to cognitive functions. Scarcity of economic resources can negatively impact cognitive functions through both physiological and psychological channels. Limited resources may lead to poor nutrition or inadequate sleep, which directly impair cognitive ability (Bond et al., 2022; **?**). Financial scarcity may also increase stress and anxiety, consuming mental bandwidth that could otherwise be used for cognitive tasks (Duquennois, 2022; Kaur et al., 2025). The cash transfer we study provides a positive shock to household resources that could improve cognitive functioning through any of these channels.

The second component connects cognitive functions to exam performance. Successful performance on standardized tests requires sustained deployment of core cognitive functions including attention, working memory, and fluid intelligence (Ackerman, 2011). These functions affect multiple aspects of test-taking. Attention determines a student's ability to maintain focus throughout a lengthy exam and process all relevant information in complex questions (Reyes, 2023; Brown et al., 2024). Working memory enables students to hold and manipulate information while solving multi-step problems (Engle, 2002; McCabe et al., 2010; Berger et al., 2025). Fluid intelligence supports the pattern recognition and abstract reasoning needed for many exam questions (Unsworth et al., 2014).

Figure B1: Conceptual framework linking monetary transfers to exam performance

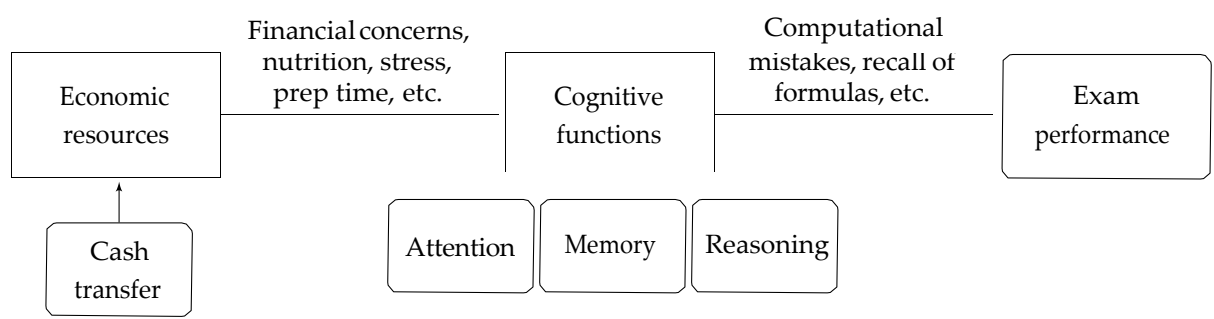